\documentclass[a4paper,11pt]{article}
\pdfoutput=1 

\usepackage{jcappub} 

\usepackage[T1]{fontenc} 

\usepackage{graphicx}

\title{Polarized Radiative Transfer of Kerr-Newman Black Hole}

\author[a]{Xin Li,}
\author[a,1]{Guo Sen,\note{Corresponding author.}}
\author[a]{Pei Wang,}
\author[b]{En-Wei Liang,}
\author[a]{Huan Deng,}
\author[c]{Yu Liang,}
\author[a]{Xiao-Xiong Zeng,}
\author[d]{Kai Lin,}
\author[e]{Qing-Quan Jiang}

\affiliation[a]{College of Physics and Electronic Engineering, Chongqing Normal University, Chongqing 401331, People's Republic of China}
\affiliation[b]{Guangxi Key Laboratory for Relativistic Astrophysics, \\ School of Physical Science and Technology, Guangxi University, Nanning 530004, People's Republic of China}
\affiliation[c]{School of Big Data and Artificial Intelligence, Fuyang University of Technology, Fuyang 236000, People's Republic of China}
\affiliation[d]{Universidade Federal de Campina Grande, Campina Grande, PB, Brasil, \\ Instituto de F\'isica, Universidade de S\~ao Paulo, S\~ao Paulo, Brasil}
\affiliation[e]{School of Physics and Astronomy, China West Normal University, Nanchong 637000, People's Republic of China}

\emailAdd{sguophys@126.com}

\abstract{In this analysis, we investigate the polarization radiation imaging of Kerr-Newman black holes, with a particular focus on the impact of black hole charge on photon propagation and polarization characteristics. By extending the traditional Walker-Penrose method, which is limited by its reliance on specific symmetric structures and Killing tensors, we overcome these limitations by constructing an ordinary differential equations (ODEs) numerical framework that combines the photon orbit equation with the polarization parallel transport equation. This allows for the self-consistent evolution of photon trajectories and polarization states in any spacetime backgrounds without relying on specific symmetries. Using this framework, we analyze the effects of black hole spin and charge on the polarization characteristics of radiation from both prograde and retrograde accretion disks. Our results show that black hole charge can significantly modify photon trajectories and polarization patterns: increasing charge compresses and distorts the EVPA structure on photon-ring scales, inducing localized rotations and asymmetries that may provide a potential diagnostic of a nonzero black hole charge.}

\begin{document}
\maketitle
\flushbottom

\section{Introduction}
\label{sec:intro}

The study of black hole images provides a powerful means to test strong field gravity and investigate the characteristics of accretion disks. The advent of the Event Horizon Telescope (EHT) has ushered in a new era of direct observational studies, with iconic images of the supermassive black holes in M87 and Sgr~A$^{*}$ offering direct visual evidence of the black hole shadow and the bright emission rings surrounding it~\cite{EventHorizonTelescope:2019dse, EventHorizonTelescope:2022wkp}. These observations enable rigorous testing of general relativity. The theoretical framework used to interpret these images is primarily based on the intensity observed, which is modeled by simulating the accretion flow around the black hole. Due to the strong gravitational lensing effect, light from the accretion flow follows highly curved trajectories, resulting in photon rings and the black hole shadow near the event horizon. These phenomena are further simulated through ray tracing codes. The light emitted by the accretion flow moves along these curved paths, and the resulting simulations yield black hole images, including the generation of photon rings and shadows near the event horizon. This theoretical framework has been shown to constrain key black hole parameters and encode the effects in the brightness distribution function, influencing the apparent size and shape of the shadows. The agreement between theoretical predictions and observational data has been notably successful. Furthermore, the black hole images in various gravitational backgrounds have been explored in several important studies~\cite{Hou:2022eev,Hou:2023bep,Zhang:2024lsf,Huang:2024wpj,Zhang:2025vyx,Meng:2025ivb,Meng:2024puu,Zeng:2020vsj,He:2024amh,Yang:2025byw,Li:2025awg,Guo:2025yin,Liang:2025bbn,Wang:2025buh,Guo:2021bhr,Hu:2024cbn,Hu:2023pyd,Zhang:2024jrw,Zeng:2025kqw,Guo:2024mij}.

Although the study of total intensity images has been groundbreaking, its results primarily encode information about spatiotemporal geometric shapes. The existence of magnetic fields around black holes is well established. Polarization observations of M87 and Sgr~A$^{*}$ reveal the characteristics of the magnetic field structure surrounding the black hole, which not only provides an image of the black hole itself but also serves as a bridge connecting strong gravitational and magnetic fields~\citep{EventHorizonTelescope:2021bee}. In theory, this crucial link is provided by polarized light, with the polarization vector directly reflecting the magnetic field structure and the orderliness of the emission region~\citep{EventHorizonTelescope:2024hpu}. According to EHT polarization observations, the electric vector exhibits a clear spiral arrangement on the emission ring, which indicates that when polarized photons pass through the strongly curved spacetime around the black hole, their paths and polarization vectors are influenced by parallel transport, resulting in a swing of the electric vector potential angle (EVPA)~\citep{EventHorizonTelescope:2021bee, EventHorizonTelescope:2024hpu}. To extract polarization information from observed images, it is necessary to solve the geodesic equation for photon trajectories in black hole spacetime and further process the parallel transport equation for polarization vectors along these photon geodesics to characterize the evolution of polarization vectors.In the context of equatorial accretion disks, Narayan $et~al$. utilized the conservation of the Walker-Penrose constant to analytically estimate the polarization of the Schwarzschild black hole image, successfully reproducing the EVPA pattern and relative polarization intensity observed by EHT for M87~\citep{EventHorizonTelescope:2021btj}. Additionally, a toy model of the equatorial radiation source of a Kerr black hole was constructed, and a radiation polarization image was obtained to distinguish the geometric effects of black hole spin on parallel photon transport from the complex coupling of gravitational and electromagnetic processes in the emission region~\citep{Gelles:2021kti}. Recent observations indicate that the total intensity and linear polarization of M87 have undergone significant variations over three cycles, and the azimuthal brightness distribution of the total intensity image changes annually~\citep{EventHorizonTelescope:2025vum}. The next-generation Very Long Baseline Interferometry (VLBI) device, the ngEHT, is expected to achieve multi-wavelength and multi-messenger observations with an angular resolution of approximately 10 $\mu$as~\citep{Lico:2023mus}. These polarization observations provide a novel method for studying black hole parameters, including the clockwise polarization vortices generated by the high spin of Kerr black holes~\citep{EventHorizonTelescope:2024rju}.

The Kerr metric has been successful in matching observational data, but it only describes the mass and spin of black holes. According to the no-hair theorem, charge is a fundamental parameter for characterizing black holes, and the rapid development of observational capabilities has made it possible to use polarization to test whether black holes carry charge~\citep{EventHorizonTelescope:2022xqj}. Consequently, the Kerr-Newman spacetime has become a key target for evaluating the effects of charge. Although astrophysical black holes are generally considered nearly neutral due to the rapid neutralization of charge by the surrounding plasma, weak charges that cannot be neglected may exist in certain accretion flow scenarios~\citep{Zajacek:2018ycb}. Observations of the supermassive black hole at the center of the Milky Way also suggest the possibility that it may carry a small amount of charge~\citep{EventHorizonTelescope:2021dqv}. When $Q=0$, the Kerr-Newman metric reduces to the Kerr metric, and when $a=0$, it reduces to the Schwarzschild metric, which is commonly used to study black hole problems. The presence of charge alters the spacetime structure of the black hole, changing the radius of the event horizon and the ergosphere, which in turn affects the evolution of the null geodesic and polarization.This effect may leave unique imprints that, in principle, could be observed. Discussions on the shadow of a Kerr-Newman black hole indicate that as the charge increases, the size of the shadow gradually decreases~\citep{Tsukamoto:2017fxq}. Hou $et~al$. studied the multi-lens effect of Kerr-Newman black holes, including numerical analysis of photon orbits and the shape of the shadow in Kerr-Newman spacetime. They further investigated the influence of charge on high-order images of black holes, showing that charge, as a black hole parameter, can induce discontinuous features in the photon ring structure~\citep{Hou:2022gge}. Additionally, we employed elliptic integration and ray-tracing methods to analyze photon trajectories around the Kerr-Newman black hole and derived images of the Kerr-Newman black hole surrounded by a thin accretion disk~\citep{Guo:2024mij}.

Current theoretical studies of black hole polarization images primarily rely on simplified emission models, which consist of three main components: black hole imaging theory, general relativistic ray tracing, and polarized radiation transport~\citep{Zhang:2021hit,Qin:2021xvx,Palumbo:2022pzj,Hu:2022sej,Liu:2022ruc,Zhang:2022klr,Delijski:2022jjj,Qin:2022kaf,Lee:2022rtg,Qin:2023nog,Deliyski:2023gik,Shi:2024bpm,Chen:2024cxi,Angelov:2025rut}. Some studies also utilize more established numerical frameworks ($\textbf{ipole}$) for discussion. These works require solving the Hamilton-Jacobi equation in four-dimensional spacetime while tracking Stokes parameters along null geodesics~\citep{Loktev:2021nhk,Emami:2022kci,Galishnikova:2023ltq}. In terms of formulation, the polarization transfer equation is expressed using the Stokes parameters $(\textit{I},\textit{Q},\textit{U},\textit{V})$, which are related to the magnetic field geometry, electron density, frequency, and the curvature of the propagation path~\citep{Huang:2024bar}. However, the magnetic field is considered independently in these studies, and it is necessary to incorporate a coupling term that represents the magnetic field into the geodesic equation. On the other hand, the development of parallel transport algorithms for calculating polarization vectors in curved spacetime has made self-consistent simulations of polarized black hole images feasible~\citep{EventHorizonTelescope:2021btj}.

In this analysis, we focus on two main issues: the impact of black hole charge on polarization, particularly in the context of Kerr-Newman black holes, where charge effects may alter the propagation and polarization characteristics of photons. On the other hand, previous analyses of polarization have largely relied on the Walker-Penrose method, which is limited by its dependence on specific symmetric structures and Killing tensors, making it applicable only to analytically solvable spacetimes and difficult to generalize. We attempt to combine the photon orbit equation with the polarization parallel transport equation to self-consistently evolve both the photon trajectory and polarization state in complex spacetime backgrounds, without relying on specific symmetries, allowing for natural extensions to general axisymmetric or non-axisymmetric spacetimes. It should be pointed out that we focus exclusively on the geometric effects of the Kerr-Newman spacetime on polarization transport in this work, and do not attempt to model charge-induced modifications of the accretion flow, magnetic-field structure, or emission microphysics.


The structure of this work is as follows: In Sec.~\ref{sec:2}, we review the Kerr-Newman spacetime geometry, derive the geodesic and polarization transport equations, and construct a unified system of first-order ordinary differential equations (ODE). Section~\ref{sec:3} establishes the initial conditions for ray tracing by introducing angle normalization, local tetrads, and photon four-momenta, which are essential for the polarized radiation transport. Section~\ref{sec:4} derive the the ODE system and image simulation, including EVPA vectors, streamline topology, and intensity polarization composites for both prograde and retrograde accretion disks. We summarize this paper in Section~\ref{sec:5}.

\section{ODE equations for photon dynamics and polarization transport in Kerr-Newman spacetime}
\label{sec:2}

In this section, we provide a brief review of the dynamics of Kerr-Newman black holes and discuss the polarization transport equations. In the Boyer-Lindquist coordinate system, the Kerr-Newman metric is expressed as follows~\citep{Bardeen:1973tla}
\begin{align}
ds^2 = &-\left( 1 - \frac{2Mr - Q^2}{\Sigma} \right) dt^2 - \frac{2a(2Mr - Q^2)}{\Sigma} \sin^2\theta \, dt \, d\phi  + \frac{\Sigma}{\Delta} dr^2 + \Sigma \, d\theta^2 \nonumber\\
&+ \left( r^2 + a^2 + \frac{a^2 (2Mr - Q^2)}{\Sigma} \sin^2\theta \right) \sin^2\theta \, d\phi^2,
\label{eq-1}
\end{align}
where
\begin{equation}
\Delta = r^2 - 2Mr + a^2 + Q^2,~~~\Sigma = r^2 + a^2 \cos^2\theta,
\label{eq-2}
\end{equation}
in which the $M$ denotes the mass of the black hole, the $a$ and $Q$ represent the spin and charge of the black hole, respectively. The condition for the existence of an event horizon is $Q^{2}+a^{2} \leq M^{2}$. It is noteworthy that as the charge $Q$ approaches zero, the Kerr-Newman black hole reduces to the Kerr black hole. Moreover, in the limit where the spin $a$ tends to zero, the metric further reduces to the Schwarzschild solution.

We describe the dynamics of photons in the Kerr-Newman geometry using the Hamiltonian formalism. The Hamiltonian is given by $\textsl{H} = \tfrac{1}{2} g^{\mu\nu} p_\mu p_\nu$, where $p_{\alpha}$ represents the covariant four-momentum of the photon. For massless particles, the trajectory is constrained by $(\textsl{H}=0)$. We restrict the photon motion to the equatorial plane $(\theta=\pi/2)$, and the covariant momentum components are given by $p_{\mu} = (p_{t},p_{r},0,p_{\phi})$, where the time and azimuthal components correspond to conserved quantities: the energy $E=-p_{t}$ and the angular momentum $L=-p_{\phi}$.The Hamiltonian can be expressed as $\textsl{H}(r, p_r; E, L_z) = g^{tt} p_t^2 + 2 g^{t\phi} p_t p_\phi + g^{\phi\phi} p_\phi^2 + g^{rr} p_r^2$. The radial effective potential is given by $R(r) = g^{rr} p_r^2 = - \left( g^{tt} E^2 + 2 g^{t\phi} E L_z + g^{\phi\phi} L_z^2 \right)$ as derived by Garnier et al~\citep{Garnier:2023lph}. Photon motion is governed by the condition $R(r) \geq 0$. The formation of photon rings occurs when both $R(r) = 0$ and $dR(r)/dr=0$, and the structure of $R(r)$ determines whether photons escape to infinity, fall into the black hole, or form stable/unstable orbits. In subsequent numerical simulations, the conserved quantities $(E,L)$ are used to construct the initial conditions and track photon trajectories in Kerr-Newman spacetime~\citep{Chen:2022ewe}.

According to the condition for circular orbits and the two conserved quantities, the covariant four-momentum at the turning point of a circular equatorial orbit can be expressed as $p_{\mu}=(-E,0,0,L)$, where the time component corresponds to the negative energy, the radial component vanishes at the turning point, the polar component is zero due to the motion being confined to the equatorial plane, and the azimuthal component represents the conserved angular momentum. In Kerr-Newman spacetime, the four-velocity of photons is given by
\begin{equation}
U^\mu = \left(\frac{dt}{d\lambda},\, \frac{dr}{d\lambda},\, \frac{d\theta}{d\lambda},\, \frac{d\phi}{d\lambda}\right),
\label{eq-3}
\end{equation}
where $\lambda$ is the affine parameter along the photon's trajectory. It is important to note that we set the photon four-velocity as the initial condition for numerical integration, which provides the initial velocity direction in the geodesic equation. Additionally, the wave vector $k^{\mu}$ of the incident photon is defined, which also serves as the boundary condition for the polarization transport equation that we will discuss subsequently.

Next, we unify the photon geodesic equation and the parallel transport equation to form a system of first-order ODEs suitable for numerical integration. The photon geodesic equation is given by
\begin{equation}
\frac{dx^\mu}{d\lambda} = k^\mu,
\qquad
\frac{dk^\mu}{d\lambda} = -\Gamma^\mu_{\;\nu\rho} k^\nu k^\rho ,
\label{eq-4}
\end{equation}
where the Christoffel symbols are defined in terms of the inverse metric as $\Gamma^\alpha_{\;\mu\nu}= \frac{1}{2} g^{\alpha s}\bigl( \partial_\mu g_{\nu s}$
\linebreak[4]
$+ \partial_\nu g_{\mu s} - \partial_s g_{\mu\nu} \bigr)$. Assuming the polarization vector is $f^{\mu}$, the parallel transport equation is expressed as
\begin{equation}
\frac{df^\mu}{d\lambda} + \Gamma^\mu_{\;\nu\rho} k^\nu f^\rho = 0,
\label{eq-5}
\end{equation}
These two equations serve as key inputs for determining the polarization characteristics of Kerr-Newman spacetime, allowing us to account for the correction to photon trajectories due to spacetime curvature under parallel transport, as well as the evolution of polarization vectors. Based on Eqs. (\ref{eq-4}) and (\ref{eq-5}), we construct a closed system of ODEs as follows: (\textit{i}) Introduce two conserved quantities associated with stationarity and axial symmetry: $p_t = -E$ and $p_\phi = L$; (\textit{ii}) Represent all dynamic variables as explicit functions of the affine parameter $\lambda$, i.e., $r \to r(\lambda), \quad \theta \to \theta(\lambda), \quad f^t \to f^t(\lambda), \quad p_r \to p_r(\lambda)$; (\textit{iii}) By combining the geodesic equation and the polarization transport equation into a coupled system, we obtain
\begin{align}
\frac{dt}{d\lambda} = k^t,~~~\frac{dr}{d\lambda} = k^r,~~~\frac{d\theta}{d\lambda} = k^\theta,~~~\frac{d\phi}{d\lambda} = k^\phi, ~~~\frac{dk^\mu}{d\lambda} = -\Gamma^\mu_{\;\nu\rho} k^\nu k^\rho,~~~\frac{df^\mu}{d\lambda} = -\Gamma^\mu_{\;\nu\rho} k^\nu f^\rho.
\label{eq-6}
\end{align}
This results in a system of 10 first-order ODEs, including 6 equations for the geodesic variables ($t^{'},r^{'},\theta^{'},\phi^{'},p_{r}^{'},p_{\theta}^{'}$) and 4 equations for the polarization components ($f^{t},f^{r},f^{\theta},f^{\phi}$). Consequently, the problem of discussing the polarization of Kerr-Newman black holes can be mathematically simplified to solving a system of closed ODEs. Given the appropriate initial conditions, the numerical solution of these ODEs yields the photon trajectories and polarization states, providing a foundation for subsequent radiative transfer calculations and polarization image synthesis.

It is worth noting that polarization transport can also be handled using the Walker-Penrose method, which leverages the hidden symmetry of Petrov D-type spacetimes to encode the evolution of polarization four-vectors into a conserved complex scalar~\citep{EventHorizonTelescope:2021btj,Gelles:2021kti}. However, its limitations arise from the reliance on specific symmetric structures and Killing tensors. Strictly speaking, it is only valid in analytically solvable spacetimes and is challenging to directly generalize to more complex geometries or physical situations. In contrast, our approach integrates the geodesic equation with the polarization parallel transport equations (Eqs.~\ref{eq-4}-\ref{eq-5}), evolving the photon four-momentum and polarization four-vector as unified dynamic variables. This method involves the self-consistent evolution of all dynamic quantities, such as $x^{\mu}$, $k^{\mu}$, and $f^{\mu}$, within a closed system of first-order ordinary differential equations. As a result, the photon orbit and polarization state are obtained simultaneously within the same numerical framework, facilitating a unified analysis of both geometric and polarization effects. Furthermore, constructing the ODE system relies solely on the spacetime metric and its Christoffel symbols, without requiring the explicit construction of the Killing tensor or a specific Penrose frame. This makes our approach a fully geometric and coordinate-space ODE method, which can be naturally extended to more general axisymmetric or non-axisymmetric spacetimes, as well as numerically constructed spacetime backgrounds. It is also applicable in situations involving plasma, scattering, or modified gravitational effects. Finally, directly handling the polarization four-vectors makes the interface between geometric constraints and radiation transfer or polarization imaging codes more direct. This is advantageous for developing a flexible and scalable numerical platform for systematic studies of the polarization characteristics of Kerr-Newman black holes. In summary, the Walker-Penrose constant can be regarded as a conserved quantity implicit in the ODE system.

Before performing ray-tracing, we must normalize the metric covariant tensor and the polarization angle to ensure that the polarization direction remains consistent during photon propagation. This requires maintaining both numerical stability and physical consistency in the ODE equations, while avoiding numerical discontinuities. We define the normalization function $\textit{F}(a_{1},a_{2})$ to map the azimuthal angle and polarization angle to the standard ranges: $a_{1} \rightarrow [0,2\pi)$ and $a_{2} \rightarrow [0,\pi)$. Specifically, $[x_{1},x_{2}] \rightarrow \textit{F}(a_{1},a_{2})$~\citep{Vincent:2023sbw}. Normalization ensures polarization reversal invariance, meaning that $\theta$ and $\theta+\pi$ describe the same direction.When $x_{1} < 0$, we take the mirror angle and rotate $x_{2}$ by $\pi$ to maintain a consistent direction, i.e., $x_{1} \rightarrow -x_{1}$, $x_{2} \rightarrow x_{2}+\pi$ and $x_{2}$ modulo $2\pi$. Therefore, the final normalized form is $[x_{1},x_{2}] = (|a_{1}|, (a_{2}+ \pi \delta_{a_{1}<0})~mod~2\pi)$, which ensures that the polarization angle remains physically consistent, allowing photons to bend or reverse along their trajectories.

\begin{figure}[t]
\centering
\includegraphics[width=0.45\textwidth]{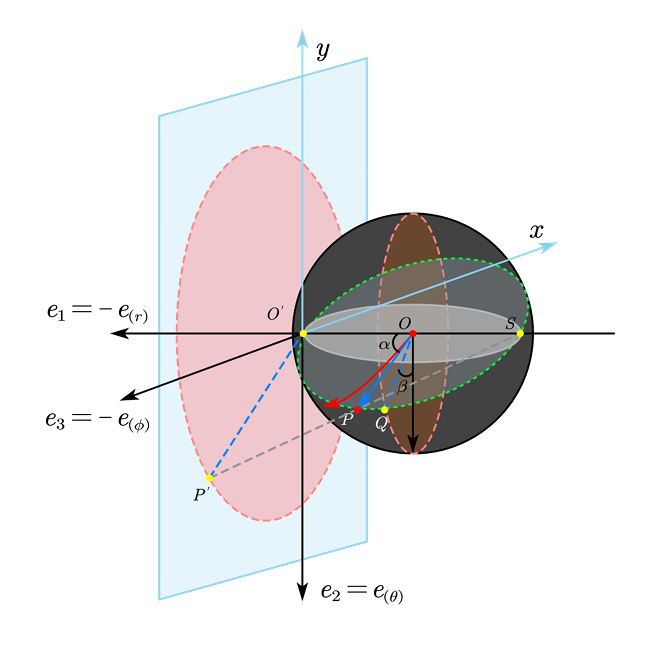}
\caption{Four-component projection method from Ref.~\citep{Zhao:2025nwi}.}
\label{fig:0}
\end{figure}

Considering the connection between ray tracing and the observer plane, it is necessary to construct a locally orthogonal tetrad in curved spacetime(Fig.~\ref{fig:0}).
Based on symmetry, we can define the physical quantities in a local observer's instantaneous rest frame, where the orthogonal basis satisfies: $\eta_{ab} = \mathrm{diag}(-1, +1, +1, +1)$ and $g_{\mu\nu} e^\mu_{(a)} e^\nu_{(b)} = \eta_{ab}$, where $\eta_{a,b}$ is the Minkowski metric~\citep{Bardeen:1972fi}. The observer's four-velocity can be expressed as
\begin{align}
&e^\mu_{(0)} = \sqrt{-\frac{g_{\phi\phi}}{g_{tt} g_{\phi\phi} - g_{t\phi}^2}} \left( 1, \, 0, \, 0, \, -\frac{g_{t\phi}}{g_{\phi\phi}} \right),~~e^\mu_{(1)} = \left( 0, \, -\frac{1}{\sqrt{g_{rr}}}, \, 0, \, 0 \right),\nonumber\\
&e^\mu_{(2)} = \left( 0, \, 0, \, \frac{1}{\sqrt{g_{\theta\theta}}}, \, 0 \right),~e^\mu_{(3)} = \left( 0, \, 0, \, 0, \, -\frac{1}{\sqrt{g_{\phi\phi}}} \right).
\label{eq-7}
\end{align}
These vectors together form the local orthogonal basis required to extract observed quantities such as photon energy, redshift, and polarization angle. Next, we associate the image plane coordinates with the normalized field-of-view (FOV) coordinates. Using the proposed fisheye camera method by Ref.~\citep{Guo:2024mij}, we map the pixels to normalized screen coordinates:
\begin{equation}
(x_{\mathrm{scr}}, y_{\mathrm{scr}}) = \frac{2 \tan(\mathrm{fov}/2)}{N_{\mathrm{pix}}} \left( i - \tfrac{1}{2}(N_{\mathrm{pix}}+1), \, j - \tfrac{1}{2}(N_{\mathrm{pix}}+1) \right),
\label{eq-8}
\end{equation}
where \(fov\) is the field-of-view angle. This mapping centers the image at $(0,0)$ and normalizes it to the $[-\tan(fov/2),\tan(fov/2)]$ range, then converts the coordinates to spherical orientation angles~\citep{Vincent:2011wz}:
\begin{equation}
\{ \theta_x, \psi_x \} = \textit{F}\left( 2 \arctan \left( \tfrac{1}{2} \sqrt{x_{\mathrm{scr}}^2 + y_{\mathrm{scr}}^2} \right), \, \arctan \tfrac{-y_{\mathrm{scr}}}{-x_{\mathrm{scr}}} \right).
\label{eq-9}
\end{equation}
We then construct a three-dimensional vector for the propagation direction in the local observer's frame, given by:
\begin{equation}
\vec{v} = \left( \cos \theta_x, \, \sin \theta_x \cos \psi_x, \, \sin \theta_x \sin \psi_x \right).
\label{eq-10}
\end{equation}
The photon's initial four-momentum is expressed as:
\begin{equation}
\lambda^\mu = -\kappa \, e^\mu_{(0)} + v^{(1)} e^\mu_{(1)} + v^{(2)} e^\mu_{(2)} + v^{(3)} e^\mu_{(3)},
\label{eq-11}
\end{equation}
where $\kappa$ is a normalization constant, $\overrightarrow{v}$ is the propagation direction vector, and $\lambda^{\mu}\lambda_{\mu}=0$ ensures that the four-momentum lies on a null geodesic.

\section{Initial conditions for solving the ODE in Kerr-Newman spacetime}
\label{sec:3}

\begin{figure*}[t]
\centering
\includegraphics[width=5cm,height=5cm]{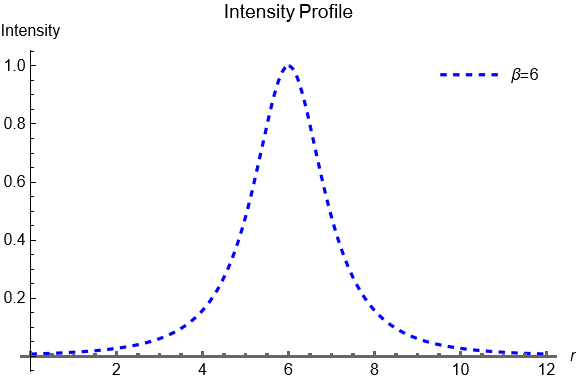}
\includegraphics[width=5cm,height=5cm]{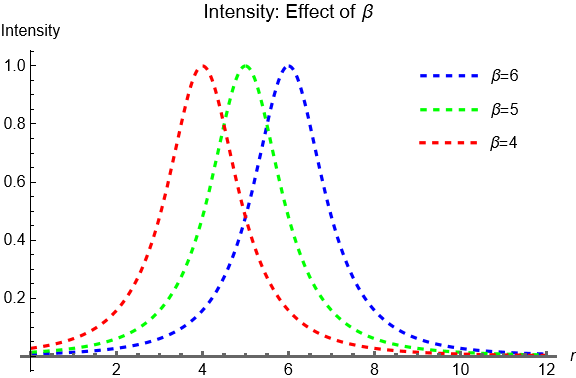}
\caption{\label{fig:1} Intensity distribution of Gaussian model. The change of parameter $\beta$ only adjusts the position where the peak appears.}
\end{figure*}

\begin{figure*}[t]
\centering
\includegraphics[width=5cm,height=5cm]{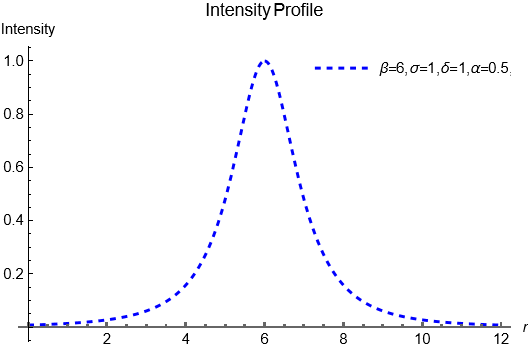}
\includegraphics[width=5cm,height=5cm]{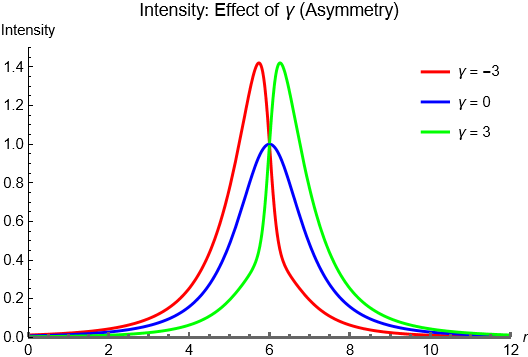}
\caption{\label{fig:2} Intensity profiles of the non-Gaussian radiation models. The variation of parameter $\gamma$ can regulate the asymmetry of the accretion disk structure.}
\end{figure*}

To solve the ODE system, it is essential to know the background emission source and the magnetic field structure that generate polarization.The source function strength, $I(r)$, sets the initial photon flux along each geodesic, providing the foundation for image synthesis~\citep{Rybicki:2004hfl}. A commonly used model for the light radiation intensity is a Gaussian like function~\citep{Kamruddin:2013iea}:
\begin{equation}
\texttt{I}[r] := \exp\!\left[-\tfrac{1}{2}\left(\gamma + \texttt{ArcSinh}\!\left(\tfrac{r-\beta}{\sigma}\right)\right)^2\right]
\big/ \sqrt{(r-\beta)^2+\sigma^2}.
\label{eq-12}
\end{equation}
In the calculation of radiative transfer in black hole images, the light intensity distribution is smooth and non-singular, with the inverse hyperbolic sine function, $\text{arsinh}$, applied at the peak to prevent numerical instability. The tail attenuation of the model is consistent with the extended characteristics of accretion disk radiation. Figure~\ref{fig:1} illustrates the smooth, symmetric emission profile of the Gaussian model, which is suitable for simplified scenarios but fails to capture the asymmetric structure of the accretion disk.
In real astrophysical environments, the radiation from accretion flows exhibits significant asymmetry and complex spatial structures. Therefore, we adopt a more flexible, non-Gaussian intensity distribution function~\citep{EventHorizonTelescope:2022wkp}:
\begin{equation}
I(r) = \Big(1 + \alpha \tanh(\gamma x)\Big) \cdot
\frac{\exp\!\left[-\tfrac{1}{2}\big(\sinh^{-1}(x)\big)^2\right]}
{(x^2 + 1)^{\delta/2}},
\qquad x = \frac{r - \beta}{\sigma}.
\label{eq-13}
\end{equation}
where $x$ is the normalized radial coordinate with scale parameter $\sigma$, and other parameter characteristics are shown in the following table. This equation introduces additional degrees of freedom, enabling it to more flexibly accommodate the asymmetry and complex structures commonly observed in real astrophysical environments. The term $\tan(\gamma x)$ introduces adjustable asymmetry while preserving the smooth peak at the center of the light source, and the behavior of the tail is governed by the power-law factor $(x^2 + 1)^{\delta/2}$. The non-Gaussian model produces a smooth peak with an extended tail and adjustable asymmetry, which can replicate observational features and provide a more realistic foundation for simulating the luminosity structure of black hole accretion systems (Figure~\ref{fig:2}).
\begin{center}
\begin{tabular}{lll}
\hline
Symbol & Physical Meaning & Default Value \\
\hline
$r$ & Radial coordinate & --- \\
$\beta$ & Position of the intensity peak & 6 \\
$\sigma$ & Width control parameter, determines the extent of the profile & 1 \\
$\gamma$ & Asymmetry control parameter & 0 \\
$\delta$ & Tail decay exponent, controls the convergence speed at the far end & 1 \\
$\alpha$ & Asymmetry amplitude & 0.5 \\
\hline
\end{tabular}
\end{center}

\begin{figure*}[t]
\centering
\includegraphics[width=5cm,height=5cm]{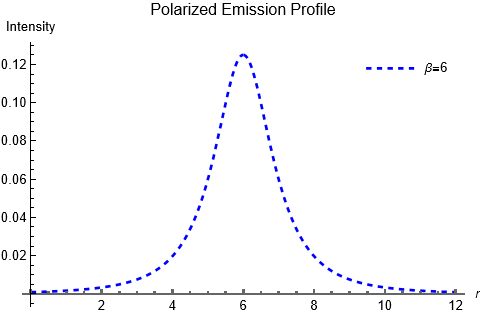}
\includegraphics[width=5cm,height=5cm]{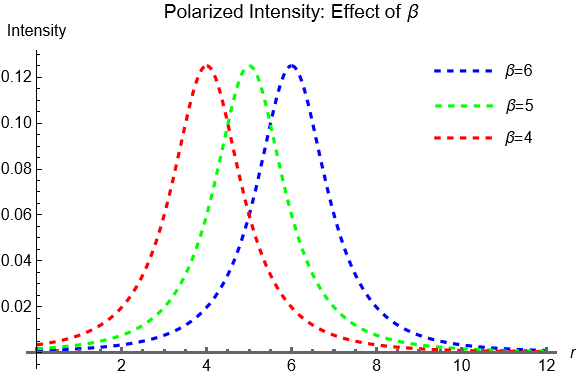}
\caption{\label{fig:3} Polarized emission profiles for this model.}
\end{figure*}

We introduce Gaussian-like functions with normalization terms to define a basic polarization emission model~\citep{Shcherbakov:2010kh}:
\begin{equation}
I_{\rm pol}(r) = \cos^2 \theta \cdot \frac{1}{2\sqrt{(r-\beta)^2 + \sigma^2}}
\exp \left[-\frac{1}{2} \left(\gamma + \sinh^{-1}\!\left(\frac{r-\beta}{\sigma}\right)\right)^2\right].
\label{eq-14}
\end{equation}
where $\beta$ sets the center position of the brightness distribution, $\sigma$ controls the smoothness and width of the radial contour, $\gamma$ introduces an offset, and $\theta$ determines the primary direction of the polarization emission process. However, this equation is idealized and lacks explicit frequency dependence, limiting its ability to capture spectral polarization coupling. Figure~\ref{fig:3} shows the polarization emission profile in this form. The left panel displays a symmetric distribution with $\beta=6$, while the right panel illustrates the effect of $\beta$ on the polarization emission position. Note that the adopted polarized emissivity model is phenomenological and is used solely to isolate geometric and transport effects; it should not be interpreted as a self-consistent description of emission around a charged black hole.In this analysis, we restrict the phenomenological model to the range of $6M \leq r \leq 25M$. For the near-horizon case ($r < 6M$), due to the complex coupling dynamics between the intrinsic electromagnetic field and the accretion flow~\cite{Hou:2024qqo}, we do not consider it here, as the treatment exceeds the scope of this study.

\begin{figure*}[t]
\centering
\includegraphics[width=3.6cm,height=3.6cm]{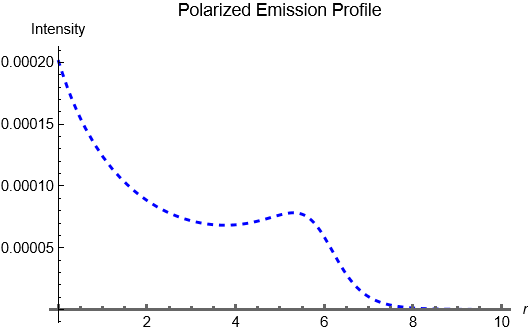}
\includegraphics[width=3.6cm,height=3.6cm]{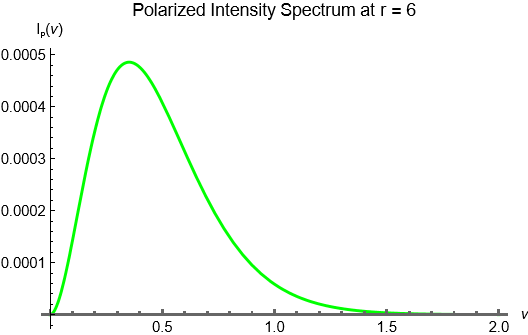}
\includegraphics[width=3.6cm,height=3.6cm]{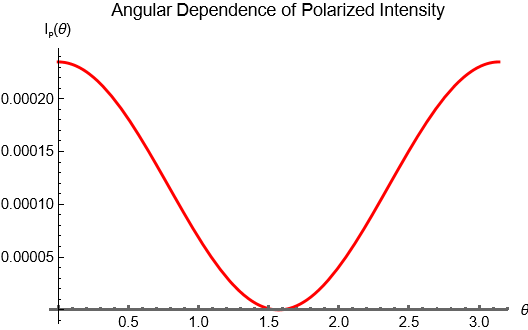}
\includegraphics[width=3.6cm,height=3.6cm]{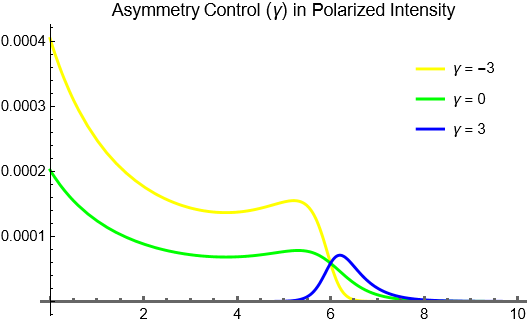}
\caption{\label{fig:4} Generalized polarized emission profiles under varying parameters.}
\end{figure*}

More generally, polarized emission depends not only on the radius $r$, but also on the viewing angle $\theta$, frequency $\nu$, and polarization degree $\Pi$. Defining the dimensionless variable $x = \frac{r - \beta}{\sigma}$, we obtain a generalized polarization source function~\citep{Moscibrodzka:2017lcu}:
\begin{equation}
I_{p}(r,\theta,\nu) = \Pi \cdot \cos^2\theta \cdot
\left( \frac{\nu^3}{e^{\nu/T(r)} - 1} \right)
\cdot \left( 1 + \alpha \tanh(\gamma x) \right)
\cdot \frac{\exp \!\left[ -\tfrac{1}{2} (\sinh^{-1}x)^2 \right]}{(x^2 + 1)^{\delta/2}},
\label{eq-15}
\end{equation}
Compared to Eq.~(\ref{eq-15}), the asymmetric parameters $(\alpha,\gamma)$ modulate the left-right symmetry of the brightness distribution, while the attenuation index $\delta$ controls the tail behavior. Figure~\ref{fig:4} demonstrates that the radial profile exhibits a spike structure with extended wings, and the angular dependence follows a $\cos^2\theta$ distribution. The asymmetric parameter $\gamma$ can be used to modulate the brightness asymmetry.

\begin{figure*}[t]
\centering
\includegraphics[width=0.23\textwidth]{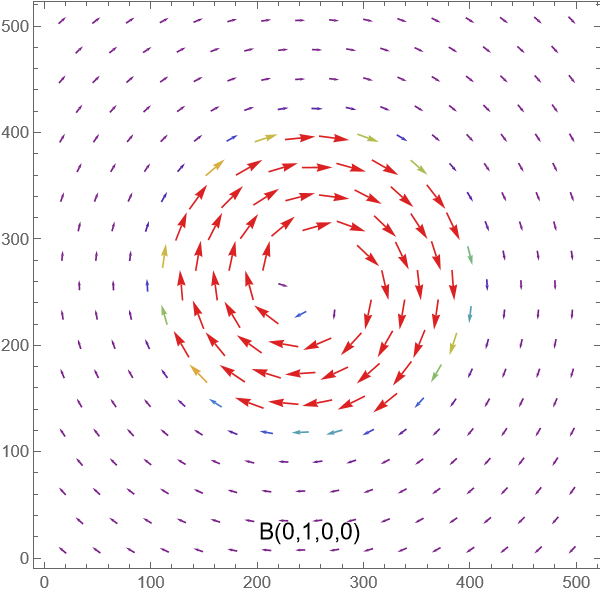}
\includegraphics[width=0.23\textwidth]{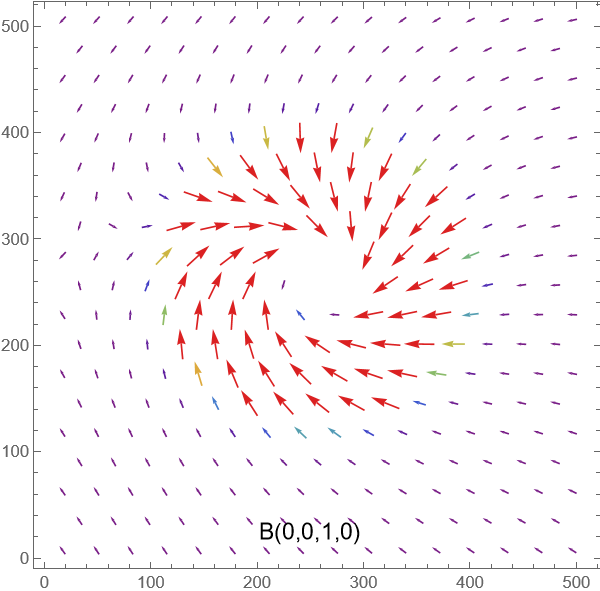}
\includegraphics[width=0.23\textwidth]{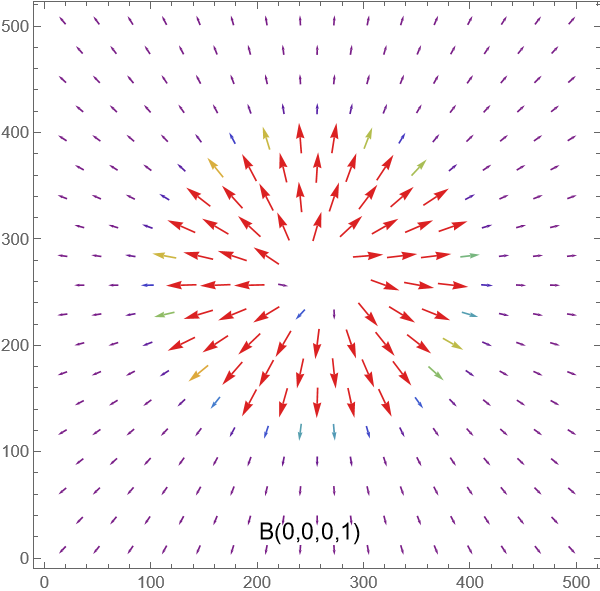}
\includegraphics[width=0.23\textwidth]{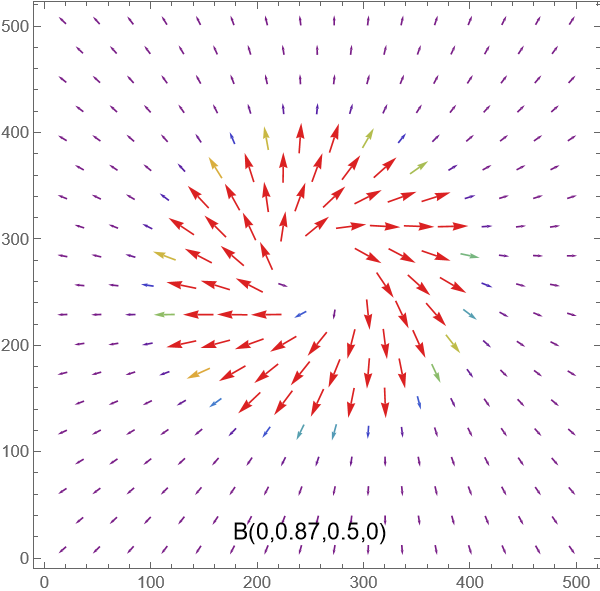}
\caption{Four magnetic field configurations.}
\label{fig:5}
\end{figure*}

In the Kerr-Newman spacetime, the magnetic field structure is represented in the locally zero reference comoving orthogonal tetrad ($e_{0},e_{1},e_{2},e_{3}$) as $B^\mu = (0,B^{1}, B^{2}, B^{3})$. In this tetrad frame, the screen-projection operator is defined as $h^\mu_{\ \nu} = \delta^\mu_{\ \nu} + e^\mu_{(0)} e_{(0)\nu}$ and $B_\perp^\mu = h^\mu_{\ \nu} B^\nu$. Thus, the local EVPA is given by $\psi = \mathrm{arg}\,(B_\perp^{(1)} + i\,B_\perp^{(2)}) + \frac{\pi}{2}$ and $\tan(2\chi) = U/Q$. When ignoring Faraday rotation, the EVPA is transmitted parallel to the imaging plane along the null geodesic. According to Refs.~\citep{Ricarte:2022sxg,Emami:2022kci,EventHorizonTelescope:2024hpu,EventHorizonTelescope:2021bee}, the pure radial field, pure polar field, pure azimuthal field, and mixed radial-polar field can be considered, as shown in Fig.~\ref{fig:5}. For pure radial fields, $B^\mu=(0,1,0,0)$, the polarization vector forms a concentric ring pattern, reflecting the $E \perp B$ projection. Near the photon ring, strong gravitational lensing and frame dragging enhance the distortion. For pure polar fields, $B^\mu=(0,0,1,0)$, the projected field tilts in the image plane, resulting in a vortex-like distortion. The degree of distortion depends on the spin of the black hole and the observer's inclination angle.For pure azimuthal fields, $B^\mu=(0,0,0,1)$, the projected $B_\perp$ is almost radial, and high-order images exhibit azimuthal periodicity. For mixed magnetic fields, $B^\mu=(0,0.87,0.5,0)$, the superposition of radial and polar components produces an open spiral EVPA pattern. The width and pitch angle of the spiral arm depend on the ratio $B^{1}/B^{2}$, which is similar to that observed in M87.

\begin{figure}[t]
\centering
\includegraphics[width=5cm,height=5cm]{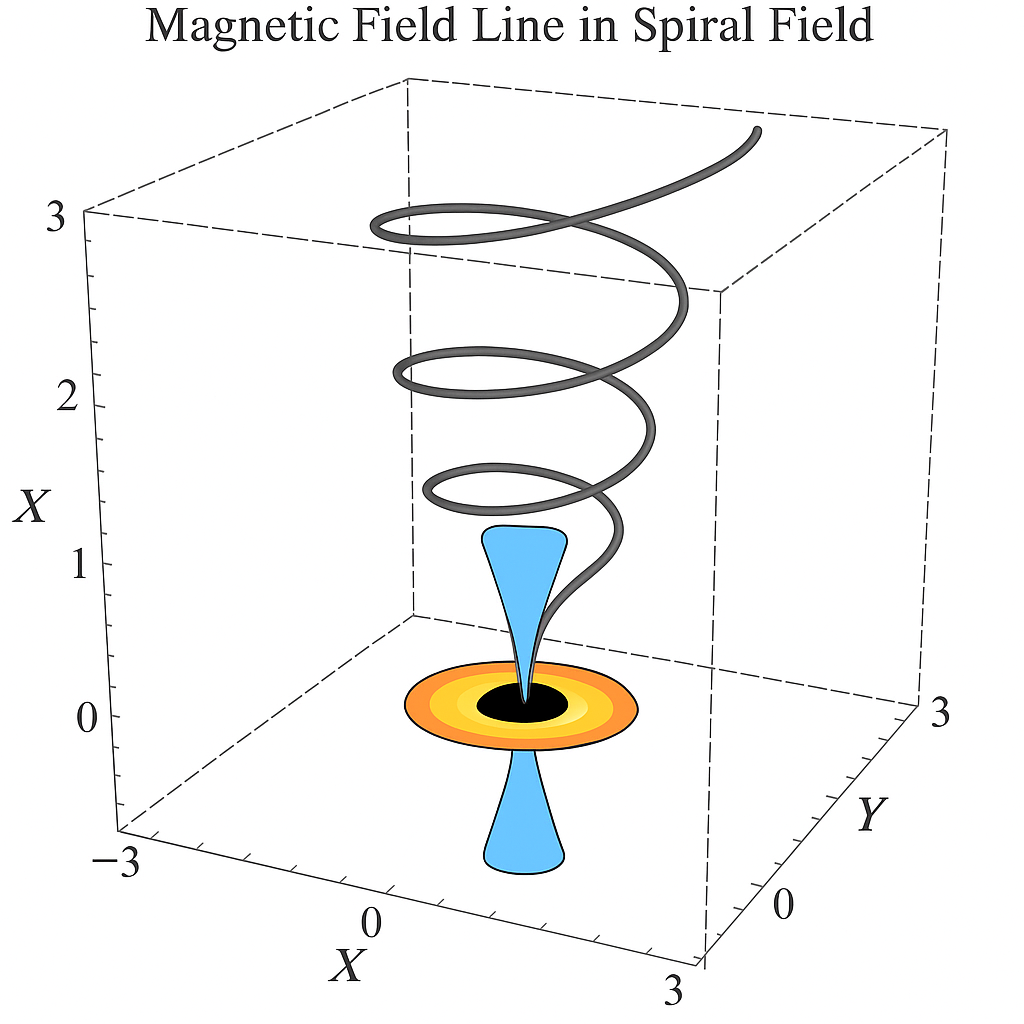}
\caption{\label{fig:6} Schematic magnetic field line in a helical configuration.}
\end{figure}

The polarization image at the field of view scale indicates that the differential rotation of the internal accretion flow rolls the large-scale polar flux into a ring-like dominant structure and adjusts the pitch angle through relativistic aberration and gravitational lensing~\citep{EventHorizonTelescope:2024hpu}. Inspired by this, a simple spiral magnetic field is employed in this analysis, which captures the basic geometric shape and facilitates the analysis of polarized radiation transmission. The magnetic field is expressed as $B^\mu(r)=(0,B^{r},B^{\theta},B^{\phi})$. Figure~\ref{fig:6} illustrates a spiral magnetic field anchored in the inner disk (orange ring) and wound by differential rotation before entering the jet (blue cone). As the field propagates to the image plane, frame dragging and gravitational lensing effects further rotate the EVPA.

\begin{figure}[t]
\centering
\includegraphics[width=5cm,height=5cm]{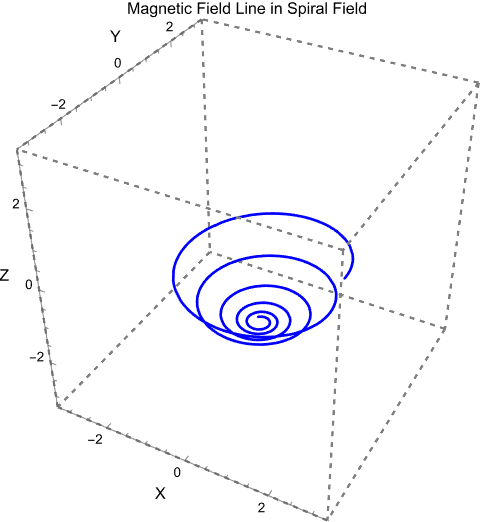}
\includegraphics[width=5cm,height=5cm]{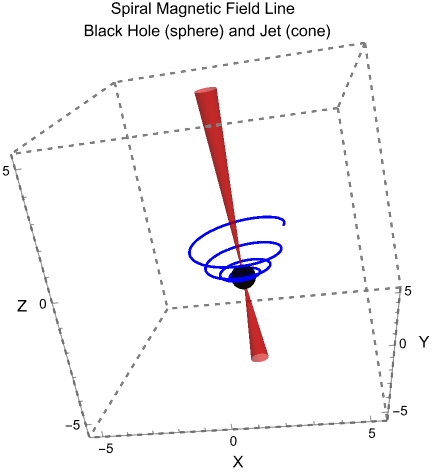}
\caption{\label{fig:7} Helical magnetic streamlines and geometry.}
\end{figure}

Note that the magnetic field in the polarization emission model is jointly influenced by the intrinsic magnetic field of the black hole and the magnetic field of the accretion flow, i.e., $B_{\rm total} = B_{\rm BH} + B_{\rm model}$~\cite{EventHorizonTelescope:2019dse,EventHorizonTelescope:2022wkp}. Firstly, we consider a circular disk or internal jet base with a moderate vertical offset that produces a circular EVPA upon projection, i.e.,
\begin{equation}
B^{r}(r)=\frac{b_{\rm ratio}}{r}, \qquad
B^{\theta}(r)=\frac{\mathrm{amp}}{r}\,\sin\!\bigl(k\ln r\bigr), \qquad
B^{\phi}(r)=\frac{1}{r}\!\left[1+\tfrac{1}{2}\cos\!\bigl(k\ln r\bigr)\right],
\label{eq-16}
\end{equation}
where $k$ is the distortion frequency of the unit $\ln r$, controlling the number of radial turns, $\mathrm{amp}$ represents the vertical fluctuation of the fixed field, and $b_{\rm ratio}$ determines the radial opening, with the annular component $B^\phi$ dominating. The local pitch angle is given by $\tan\psi(r) = \frac{\sqrt{[B^{r}(r)]^{2} + [B^{\theta}(r)]^{2}}}{B^{\phi}(r)}$, so increasing $b_{\rm ratio}$ expands the helix, while increasing $k$ increases the number of windings without altering the opening angle. Figure~\ref{fig:7} shows representative magnetic streamlines, with the left panel illustrating the case of a fixed $(k, \mathrm{amp})$ and $b_{\rm ratio} > 0$, which generates large pitch angles and rapid radial expansion. The right panel displays a schematic diagram of a black hole system with a spiral magnetic field.

In consideration of the intrinsic electromagnetic field of the Kerr-Newman black hole, its four-dimensional electromagnetic potential is given by~\cite{Kerr:1963ud}
\begin{equation}
\label{eq-17-1}
A_{\mu} {\rm d} x^{\mu} = \frac{1}{4 \pi \epsilon_{0} c}\Big (\frac{Q}{r} {\rm d}t - \frac{a c^{-1}}{r} \sin^2 \theta {\rm d} \phi\Big) / \rho^{2},
\end{equation}
where
\begin{equation}
\label{eq-17-2}
\rho^2 = r^2 + a^2 \cos^2 \theta.
\end{equation}
In the far-field approximation, Eq.~({\ref{eq-17-2}}) becomes
\begin{equation}
\label{eq-17-3}
\rho^2 = r^2 + a^2 \cos^2 \theta \approx r^2,
\end{equation}
and
\begin{equation}
\label{eq-17-4}
\frac{r}{\rho^2} = \frac{r}{r^2 + a^2 \cos^2 \theta} = \frac{1}{r} \Big(1 + \frac{a^2 \cos^2 \theta}{r^2}\Big)^{-1} \approx \frac{1}{r}.
\end{equation}
Thus, the four-potential can be approximated as
\begin{equation}
\label{eq-17-5}
A_{\mu} {\rm d} x^{\mu} \approx \frac{1}{4 \pi \epsilon_0 c} \Big(\frac{Q}{r} {\rm d} t - \frac{Q a \sin^2 \theta}{r c^2} {\rm d} \phi \Big).
\end{equation}
Hence, the components of the four-potential in the $r$-direction, $\theta$-direction, and $\phi$-direction are obtained
\begin{equation}
\label{eq-17-6}
A_{\rm r} = 0, A_{\theta} = 0, A_{\phi} \approx \frac{-1}{4 \pi \epsilon_{0} c^{2}} \frac{Q a \sin \theta}{r^2}.
\end{equation}
In spherical coordinates, the magnetic field $\overrightarrow{B}$ is given by $\nabla \times \overrightarrow{A}$
\begin{eqnarray}
\label{eq-17-6}
&&(\nabla \times \overrightarrow{A})_r = \frac{1}{r \sin \theta} \Big(\frac{\partial}{\partial \theta} (\sin \theta A_{\phi}) - \frac{\partial A_\theta}{\partial \phi}\Big),\nonumber\\
&&(\nabla \times \overrightarrow{A})_\theta = \frac{1}{r} \Big(\frac{1}{\sin \theta} \frac{\partial A_r}{\partial \phi} - \frac{\partial}{\partial r} (r A_\phi)\Big),\nonumber\\
&&(\nabla \times \overrightarrow{A})_\phi = \frac{1}{r} \Big(\frac{\partial}{\partial r} (r A_\theta) - \frac{\partial A_r}{\partial \theta}\Big).
\end{eqnarray}
Substituting $A_{\rm r} = 0, A_{\rm \theta} = 0$ and $A_{\rm \phi} = A_{\rm \hat{\phi}}$, we obtain
\begin{eqnarray}
\label{eq-17-7}
&&B_r = \frac{1}{r \sin \theta} \frac{\partial}{\partial \theta} \Big(\sin \theta A_{\rm \hat{\phi}}\Big),\nonumber\\
&&B_\theta = -\frac{1}{r} \frac{\partial}{\partial r} \Big(r A_{\rm \hat{\phi}}\Big),~~~B_\phi = 0.
\end{eqnarray}
Substituting $A_{\rm \hat{\phi}}$ yields,
\begin{equation}
\label{eq-17-8}
B_r \propto \frac{Q a \cos \theta}{r^3},~~B_\theta \propto \frac{Q a \sin \theta}{r^3},~~B_\phi = 0.
\end{equation}
Thus, the intrinsic magnetic field of the Kerr-Newman black hole takes on a dipolar form in the far region studied, and decays as higher powers of the radius. Furthermore, the components of the intrinsic magnetic field in the $r$, $\theta$, and $\phi$ directions are much smaller in magnitude compared to the corresponding components of the phenomenological helical magnetic field model introduced in this analysis, within the emission region of interest. In the present study, to highlight the effects of the Kerr-Newman spacetime geometry on polarization radiation transfer, while avoiding the complex coupling between the black hole's intrinsic electromagnetic field and the accretion flow's magnetic field, we neglect the impact of the Kerr-Newman black hole's intrinsic magnetic field on the polarization emission model.

In the equatorial direction, the projection field satisfies $(B_x,B_y)\approx(B^\phi,B^\theta)$, so the local EVPA $\chi$ is approximately given by $\chi(r) \approx \tfrac{1}{2}\arctan\!\left(\frac{B_y}{B_x}\right) + \tfrac{\pi}{2} = \tfrac{1}{2}\arctan\!\left(\frac{B^{\theta}(r)}{B^{\phi}(r)}\right) + \tfrac{\pi}{2}$. The azimuthal angle of $\chi(r)$ is directly tracked as the radius changes. In the region where the annular component dominates, the EVPA exhibits a radial dependence, while $B^{\theta}$ and $B^r$ contribute to the pitch. Characteristic sign flipping ($Q/U$) and EVPA vortices arise near the photon ring, both of which become more pronounced with increasing spin and tilt of the black hole.

The ODE equations and the initial polarization states of polarized radiation have been constructed. When the positions $x^\mu$ and photon four-momentum $k^\mu$ ($k^\mu k_\mu = 0$) are determined, the polarization four-vector $f^\mu$ must satisfy two conditions: the transversality condition $k_\mu f^\mu = 0$ of the photon momentum and the normalization condition $f^\mu f_\mu = \pm 1$. Let $U^\mu$ represent the observer's four-velocity, and $B^\mu$ represent the electromagnetic four-vector of the magnetic field. In optically thin synchrotron radiation, the EVPA is orthogonal to the projected magnetic field and can be defined using the Levi-Civita tensor covariant $f^\mu$, i.e., $f^\mu \propto \varepsilon^{\mu\nu\rho\sigma} U_\nu k_\rho B_\sigma$. Furthermore, the condition $U_\mu f^\mu = 0$ implies that $f^\mu$ lies entirely within the observer's instantaneous space. In the observer's screen, $f^\mu$ can be written as
\begin{equation}
f^\mu \;=\;
\frac{\varepsilon^{\mu\nu\rho\sigma}U_\nu k_\rho B_\sigma}
{\sqrt{\bigl|\bigl(\varepsilon^{\alpha\beta\gamma\delta}U_\beta k_\gamma B_\delta\bigr)
g_{\alpha\lambda}
\bigl(\varepsilon^{\lambda\eta\kappa\zeta}U_\eta k_\kappa B_\zeta\bigr)\bigr|}} \, .
\label{eq-17}
\end{equation}
This definition is covariant and independent of coordinate choice. In a flat spatiotemporal region, $f^\mu$ reduces to a spatial unit vector orthogonal to the plane formed by $\mathbf{k}$ and $\mathbf{B}$. Polarization is transmitted parallel to the null geodesic line to the distant observer, and $(e_{1}, e_{2})$ is projected onto the image plane as $f^{(i)} = f^\mu e^{(i)}_{\ \mu}$, yielding $\chi = \frac{1}{2} \arctan 2(f^{(2)}, f^{(1)})$. The corresponding Stokes parameters are given by
\begin{equation}
Q = I \cos 2\chi, \quad U = I \sin 2\chi,
\label{eq-18}
\end{equation}
where $I$ represents the total intensity of linear polarization. Multiple imaging can further induce characteristic sign changes in $Q/U$, and, according to Eq.~(\ref{eq-16}), the complete set of covariant polarization transfer equations along the geodesic evolution can be obtained for $(I, Q, U, V)$.

\section{Solving the ODE and image simulation}
\label{sec:4}

Using infinite photon energy as the normalized unit, the redshift factor is given by $g=\nu_{obs}/\nu_{em}$. According to Liouville's theorem, the invariant $I_\nu / \nu^3$ is conserved, and the emission from the local source model transforms as
\begin{equation}
I = g^3\, j_I(r_{\rm em}), \qquad
I_p = g^3\, j_p(r_{\rm em}),
\label{eq-19}
\end{equation}
where $j_I$ and $j_p$ represent the total emissivity and the linearly polarized emissivity in the emitter frame, respectively. The set $(x^\mu, p_\mu, f^\mu)$ is parallelly transmitted along affine parameters and propagated backward to distant observers, preserving $k_\mu f^\mu = 0$ and $U_\mu f^\mu = 0$. Using an image plane orthogonal quadrant, the transmitted polarization is projected as
\begin{equation}
E_x \propto -f_\mu\, e^\mu_{(x)},~~~E_y \propto -f_\mu\, e^\mu_{(y)}.
\label{eq-20}
\end{equation}
The linear Stokes parameters are given by
\begin{equation}
Q = I_p \cos(2\chi) = I_p\,\frac{E_x^{\,2}-E_y^{\,2}}{E_x^{\,2}+E_y^{\,2}},~~~U = I_p \sin(2\chi) = I_p\,\frac{2E_xE_y}{E_x^{\,2}+E_y^{\,2}}.
\label{eq-21}
\end{equation}
Since we are considering synchrotron radiation polarization, we set $V = 0$. Generating an image requires solving the geodesic polarization ODE system for each pixel, which involves a total of $\mathcal{O}(N^2)$ rays. The parallel structure allows for static domain decomposition in the $P$ domain, yielding
\begin{equation}
S(P) \simeq \frac{N^2\,T_{\rm ray}}{N^2\,T_{\rm ray}/P + T_{\rm oh}},
\label{eq-22}
\end{equation}
where $T_{\rm ray}$ is the average computational cost of each ray, and each geodesic is integrated with a deterministic step size and spatiotemporal structure. The integration terminates when it reaches the outer boundary, near the event horizon, or when it encounters a specified set of geometric events. At $r_{\rm em}$, parallel transmission is performed to the observation plane using Eq.~(\ref{eq-19}) and the initialized polarization vector $f^\mu_0$, with projection carried out according to Eqs.~(\ref{eq-8})-(\ref{eq-11}). The output of each ray is assembled into a two-dimensional array representing $I$, $Q$, $U$, and $V$, from which we can calculate the degree of polarization, i.e.,
\begin{equation}
{\rm DOP}(x,y)=\frac{\sqrt{Q^2+U^2}}{I},
\label{eq-23}
\end{equation}
as well as the EVPA field $\chi(x,y)$. Convergence is verified by doubling the resolution $N \to 2N$ and tightening the integration tolerances, requiring
\begin{equation}
\frac{\|I_{2N}-I_N\|_2}{\|I_{2N}\|_2}<10^{-3}, \qquad
\|\chi_{2N}-\chi_N\|_\infty < 0.5^\circ.
\label{eq-24}
\end{equation}

Ray tracing emitted at the field of view scale typically produces images with a certain dynamic range, where bright photon rings and shadows coexist, with the shadows being several orders of magnitude dimmer than the photon rings. To address this issue, pixel intensity compression maps $I \mapsto \tilde{I} \in [0,1]$ can be applied for visualization purposes. In our results, both radiative transfer and Stokes analysis use physically specific intensities, namely $I_{\max} \equiv \max_{i,j} I(i,j)$ and $\hat{I} \equiv I / I_{\max}$. Following Ref.~\citep{EventHorizonTelescope:2019dse}, we chose the following normalization:
\begin{equation}
\label{eq-25}
I=\left[\frac{\operatorname{asinh}(\alpha I)}{\operatorname{asinh}(\alpha I_{\max})}\right]^{\gamma} , \quad \alpha \!\in\! [2,5],\; \gamma \!\in\! [0.6,0.8].
\end{equation}
This function exhibits smooth and monotonic characteristics, making it particularly suitable for superimposing polarization vectors.

\begin{figure*}[t]
\centering
\includegraphics[width=5cm,height=5cm]{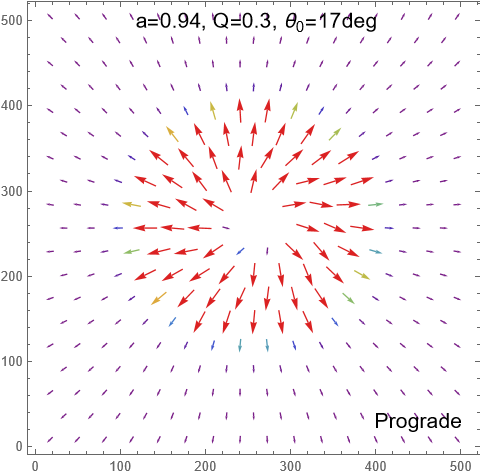}
\includegraphics[width=5cm,height=5cm]{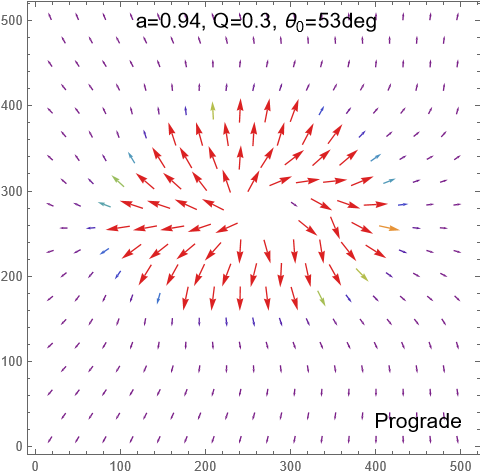}
\includegraphics[width=5cm,height=5cm]{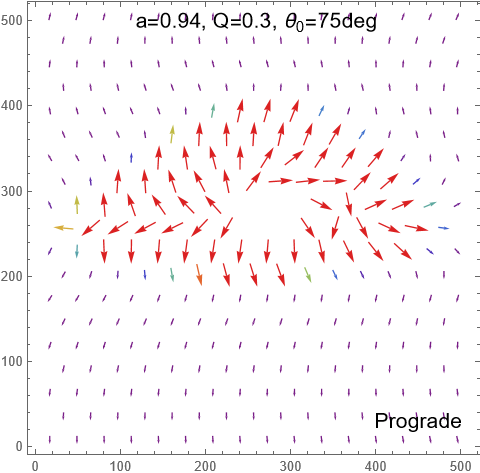}
\includegraphics[width=5cm,height=5cm]{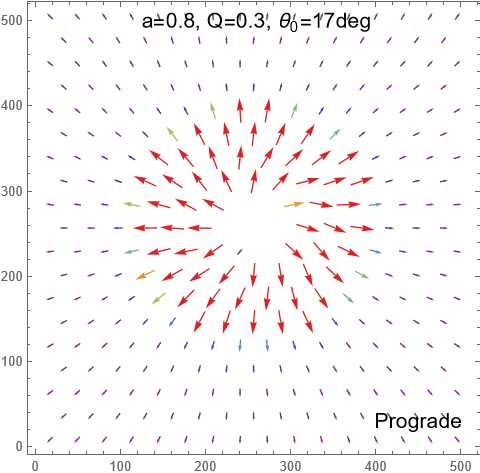}
\includegraphics[width=5cm,height=5cm]{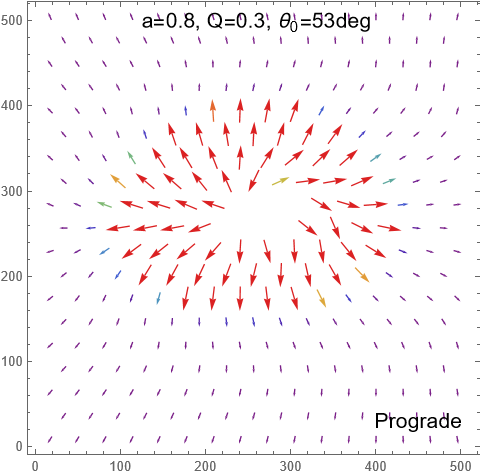}
\includegraphics[width=5cm,height=5cm]{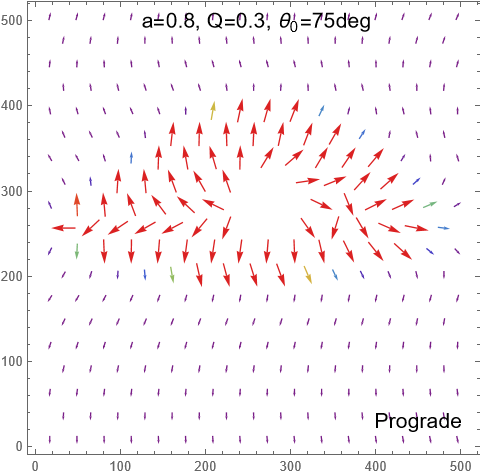}
\includegraphics[width=5cm,height=5cm]{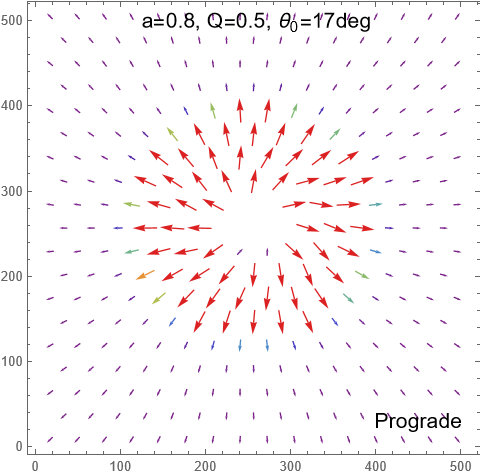}
\includegraphics[width=5cm,height=5cm]{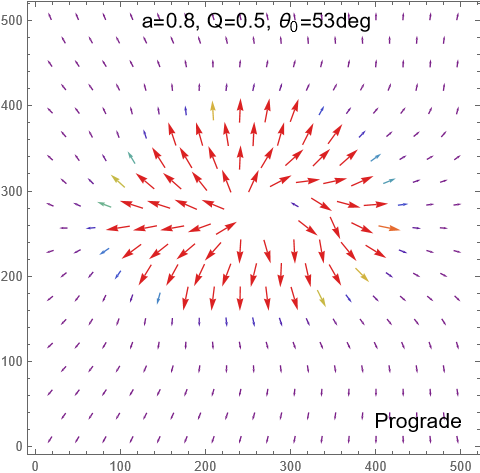}
\includegraphics[width=5cm,height=5cm]{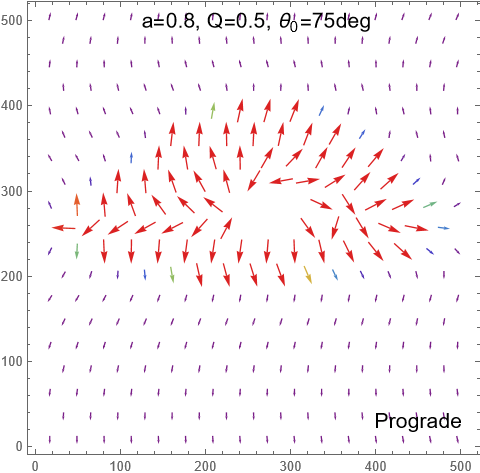}
\caption{\label{fig:8} The synchrotron radiation polarization structure surrounded by a prograde accretion disk around a Kerr-Newman black hole under different parameters.}
\end{figure*}

Figure~\ref{fig:8} shows the polarization morphology of a Kerr-Newman black hole surrounded by a prograde accretion disk, with observer inclinations of $\theta_0=17^\circ,\,53^\circ,\,75^\circ$ from left to right. At low observation angles, the polarization structure of the inner layer of the accretion disk is displayed, where the magnetic field lines are almost parallel to the disk surface, resulting in a clear symmetry of the polarization vector. As the observation angle increases, the magnetic field lines become more curved, and the spatial distribution of the polarization intensity exhibits strong asymmetry. The influence of black hole spin is primarily reflected in the geometry of the accretion disk and the magnetic field distribution. Higher spin brings the innermost edge of the accretion disk closer to the event horizon, causing the magnetic field lines to become more distorted and concentrated. With increasing relativistic effects, the polarization directions of the magnetic fields and radiation exhibit complex distortions and local rotations near the black hole. In contrast, lower spin results in a more expanded accretion disk, a more uniform magnetic field distribution, gentler changes in the polarization vector, and a simpler overall synchrotron radiation polarization structure. It is worth noting that black hole charges play a crucial role in the polarization characteristics of synchrotron radiation. The presence of electric charge not only affects the electromagnetic field configuration around the black hole but also impacts the magnetic fields in the accretion disk and the black hole polar region. When a black hole has an electric charge, the electromagnetic field becomes more complex, leading to significant changes in the bending of magnetic field lines and the distribution of polarization vectors. The influence of charge on the magnetic field is especially noticeable in the black hole polar region, where the polarization mode may exhibit stronger asymmetry at higher charges. Additionally, the electromagnetic effects caused by the charge can lead to changes in the polarization intensity and direction of synchrotron radiation, resulting in more complex and irregular polarization patterns compared to the case without charge.

\begin{figure*}[t]
\centering
\includegraphics[width=5cm,height=5cm]{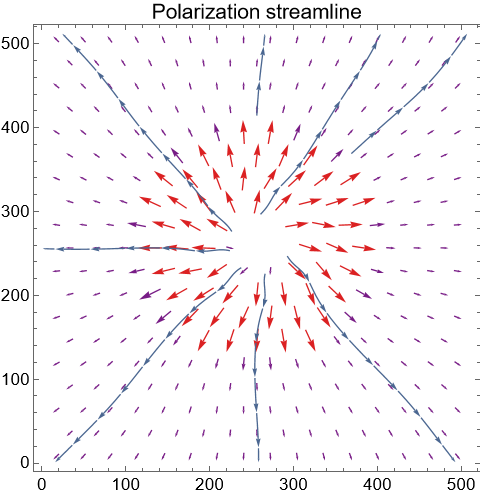}
\includegraphics[width=5cm,height=5cm]{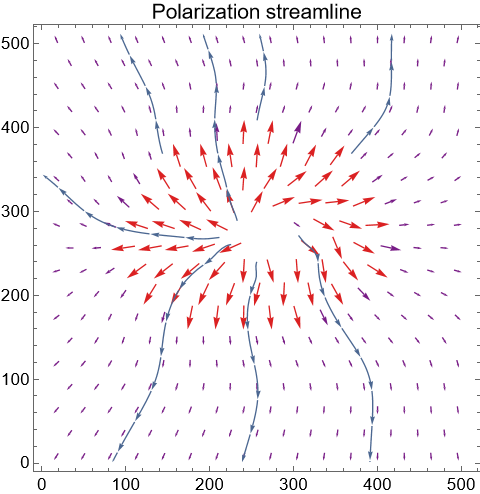}
\includegraphics[width=5cm,height=5cm]{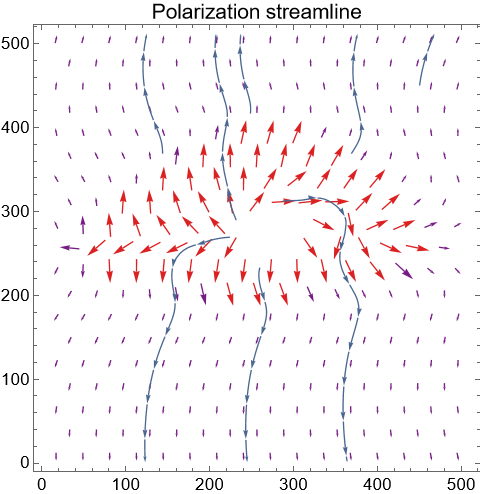}
\includegraphics[width=5cm,height=5cm]{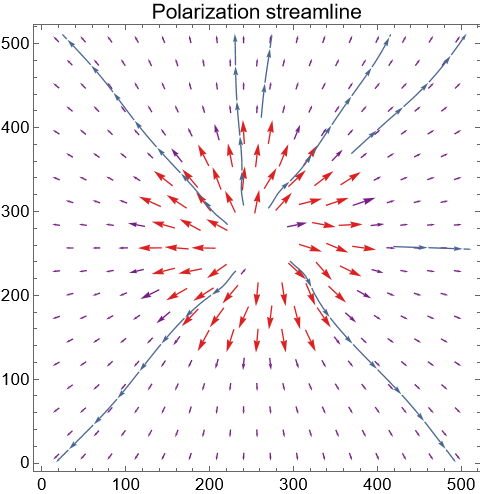}
\includegraphics[width=5cm,height=5cm]{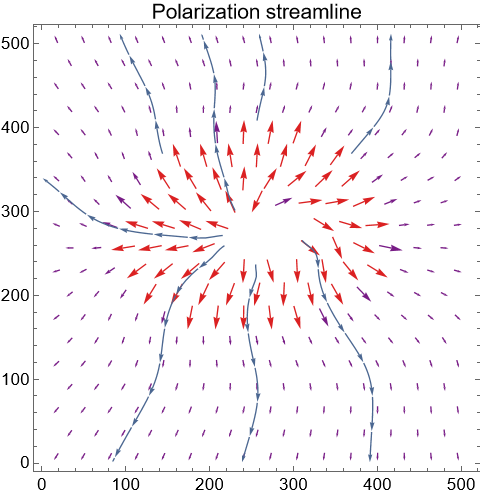}
\includegraphics[width=5cm,height=5cm]{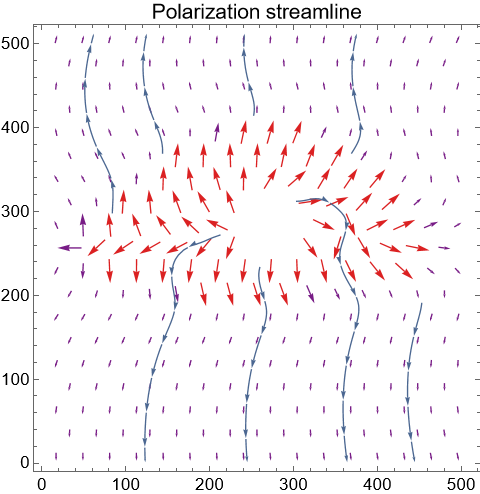}
\includegraphics[width=5cm,height=5cm]{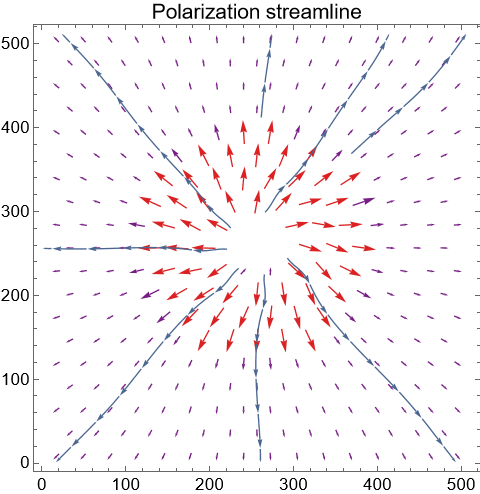}
\includegraphics[width=5cm,height=5cm]{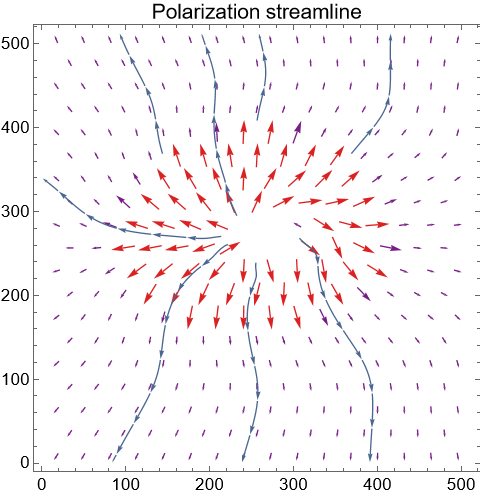}
\includegraphics[width=5cm,height=5cm]{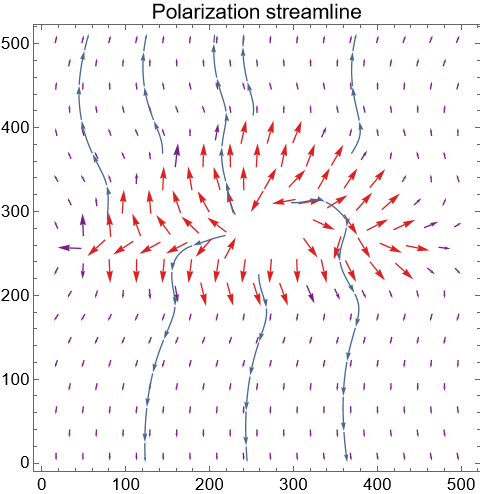}
\caption{\label{fig:9} The synchrotron radiation linearly polarized streamlines surrounded by a prograde accretion disk around a Kerr-Newman black hole under different parameters.}
\end{figure*}

Figure~\ref{fig:9} shows the distribution of synchrotron radiation linearly polarized streamlines. When observed from a smaller inclination angle, the polarization streamlines exhibit a uniform and symmetrical structure, indicating a more concentrated distribution of magnetic field and synchrotron radiation around the black hole. As the observation angle increases, the magnetic field structure inside and outside the accretion disk becomes progressively evident, and the polarization streamlines begin to show increasing distortion. At higher viewing angles, the polarization streamlines exhibit strong bending or asymmetry. Higher spin leads to significant distortion in the polarization streamlines of synchrotron radiation. In the absence of charge, the distribution of polarized streamlines is primarily governed by the spin of the black hole and the structure of the accretion disk, with the streamlines remaining relatively symmetrical. However, when the black hole charge increases, the polarization streamlines exhibit a more asymmetric structure.

\begin{figure*}[t]
\centering
\includegraphics[width=5cm,height=5cm]{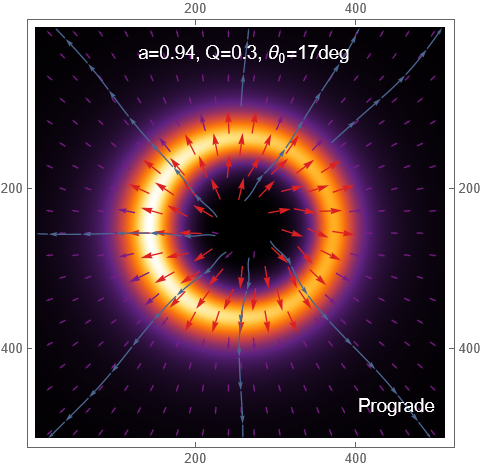}
\includegraphics[width=5cm,height=5cm]{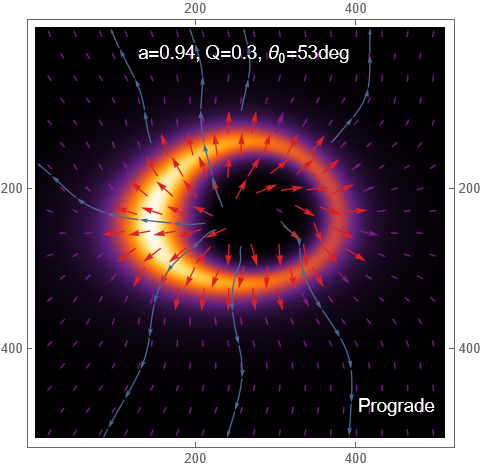}
\includegraphics[width=5cm,height=5cm]{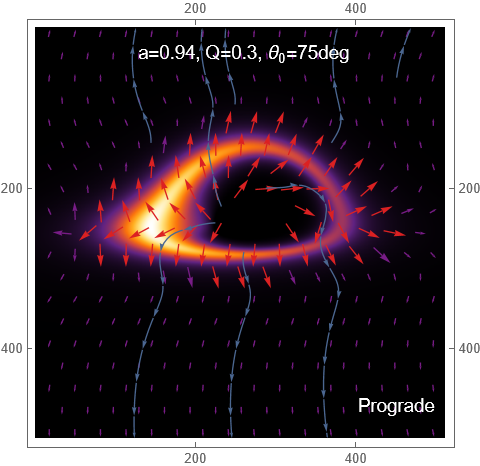}
\includegraphics[width=5cm,height=5cm]{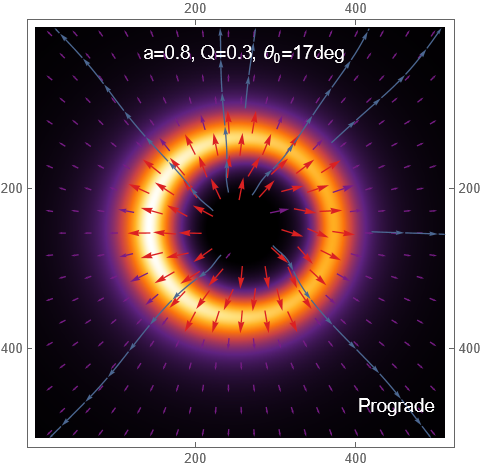}
\includegraphics[width=5cm,height=5cm]{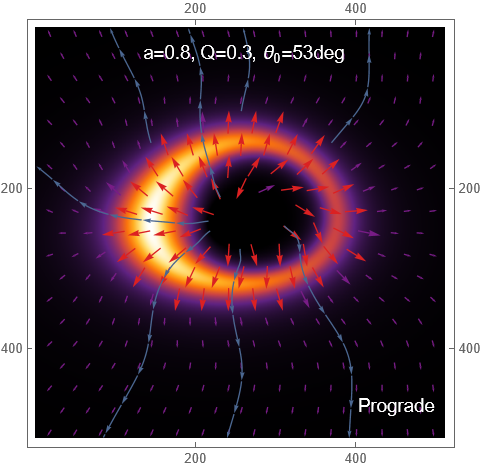}
\includegraphics[width=5cm,height=5cm]{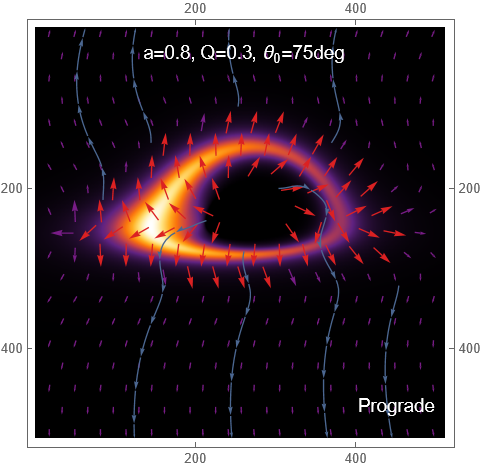}
\includegraphics[width=5cm,height=5cm]{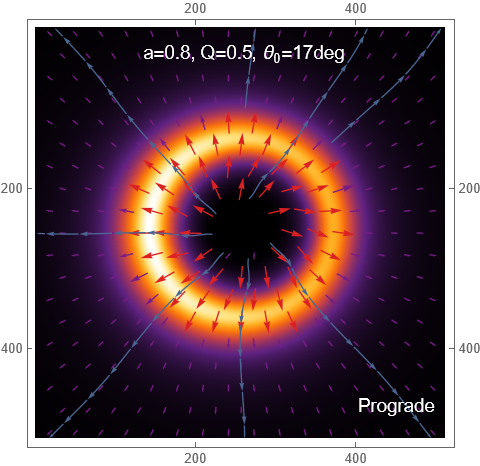}
\includegraphics[width=5cm,height=5cm]{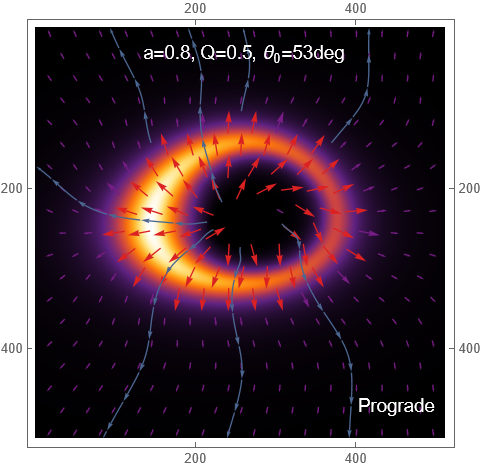}
\includegraphics[width=5cm,height=5cm]{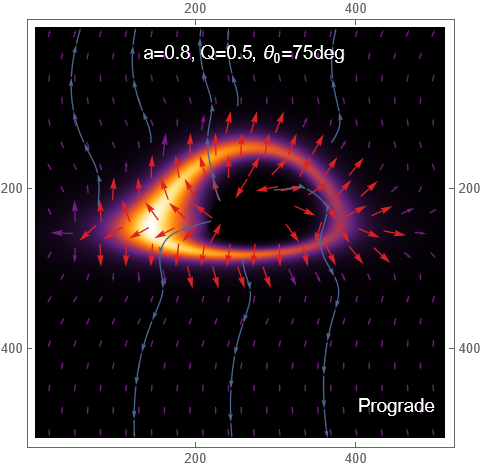}
\caption{\label{fig:10} Synchrotron radiation brightness and polarization distribution of Kerr-Newman black hole on prograde disk.}
\end{figure*}

We combine polarized images with black hole images. Figure~\ref{fig:10} shows that at higher spins, the outer part of the accretion disk is relatively smaller, and the brightness distribution of the disk exhibits a more concentrated and compact structure. At lower spins, the brightness distribution is smoother and more extended, indicating that the radiation energy is spread over a larger area. The observed tilt angle yields the same result as reported in Ref.~\citep{Guo:2024mij}. As the charge increases, the brightness of the accretion disk undergoes significant distortion in the polar and outer regions of the black hole, reflecting the asymmetry in the electromagnetic field induced by the charge. At high charge levels, the boundaries between the polar regions and radiation bands in the image become more irregular, further indicating the influence of charge on the structure of the accretion disk.

\begin{figure*}[t]
\centering
\includegraphics[width=5cm,height=5cm]{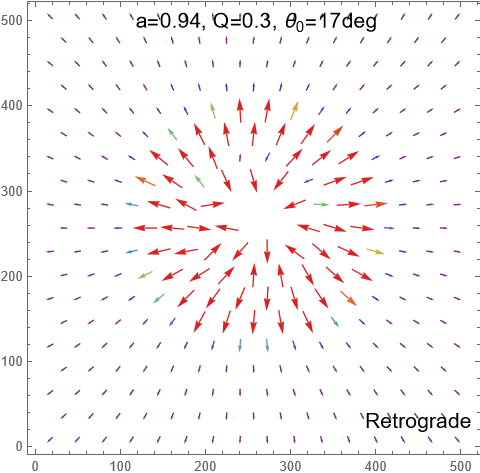}
\includegraphics[width=5cm,height=5cm]{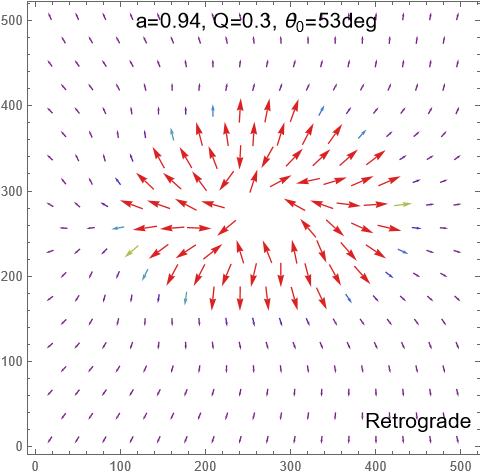}
\includegraphics[width=5cm,height=5cm]{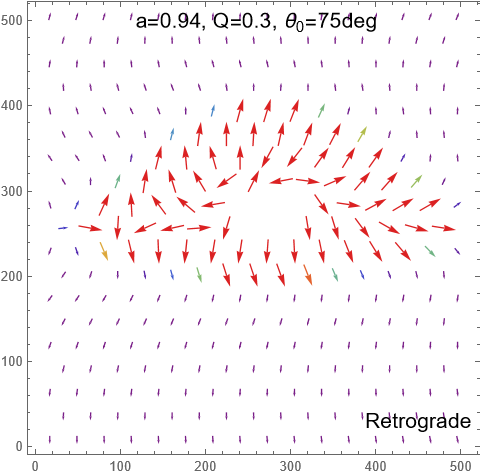}
\includegraphics[width=5cm,height=5cm]{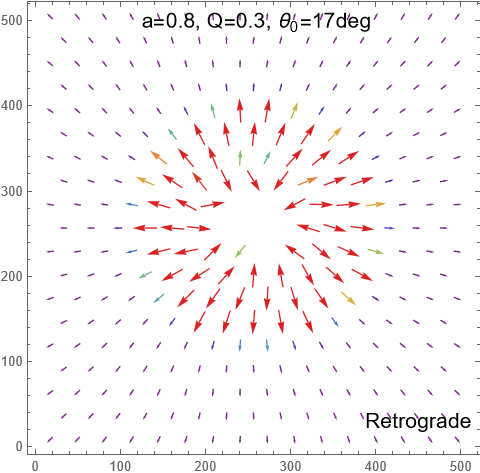}
\includegraphics[width=5cm,height=5cm]{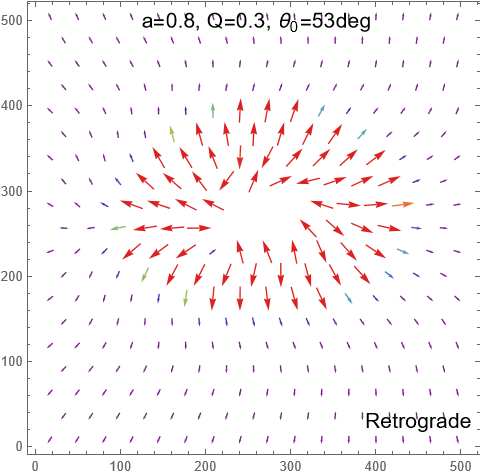}
\includegraphics[width=5cm,height=5cm]{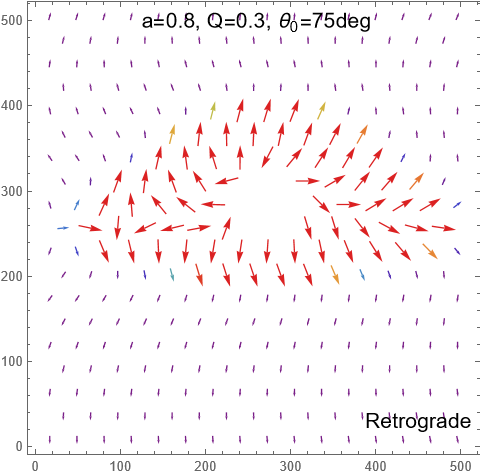}
\includegraphics[width=5cm,height=5cm]{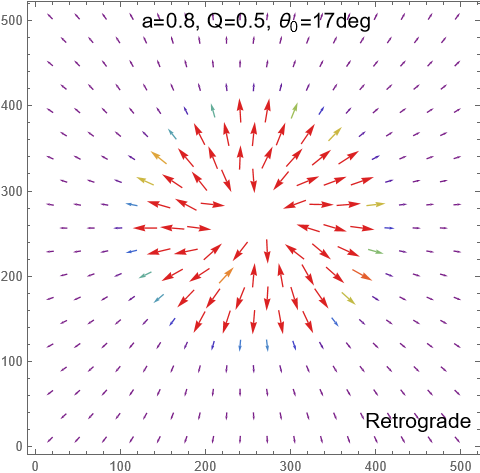}
\includegraphics[width=5cm,height=5cm]{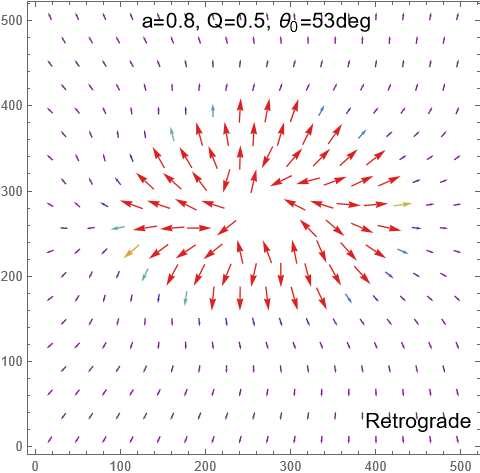}
\includegraphics[width=5cm,height=5cm]{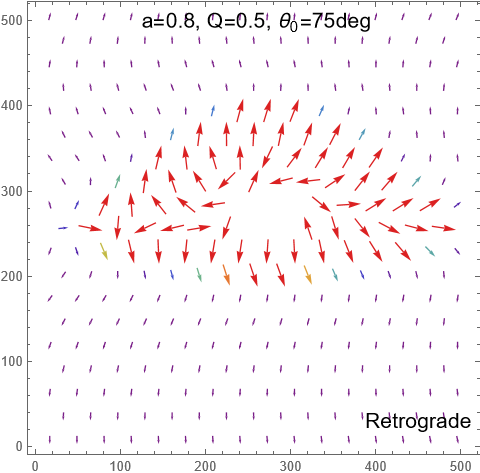}
\caption{\label{fig:11} The synchrotron radiation polarization structure surrounded by a retrograde accretion disk around a Kerr-Newman black hole under different parameters.}
\end{figure*}

\begin{figure*}[t]
\centering
\includegraphics[width=5cm,height=5cm]{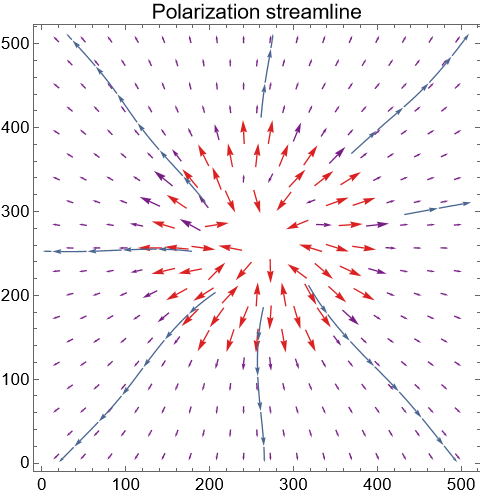}
\includegraphics[width=5cm,height=5cm]{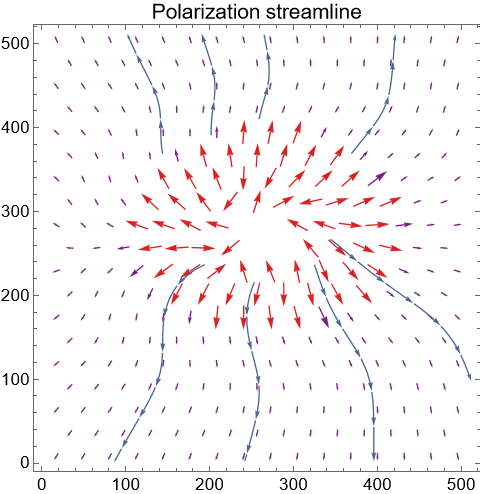}
\includegraphics[width=5cm,height=5cm]{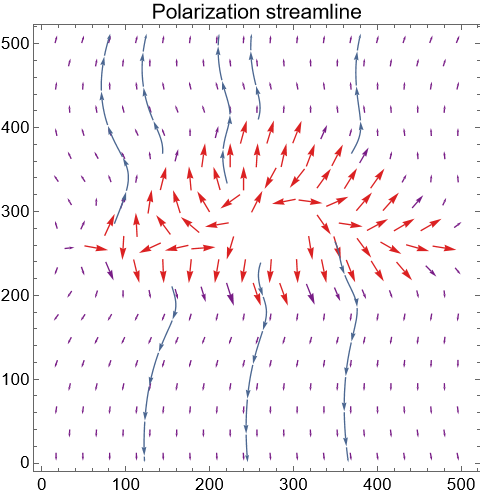}
\includegraphics[width=5cm,height=5cm]{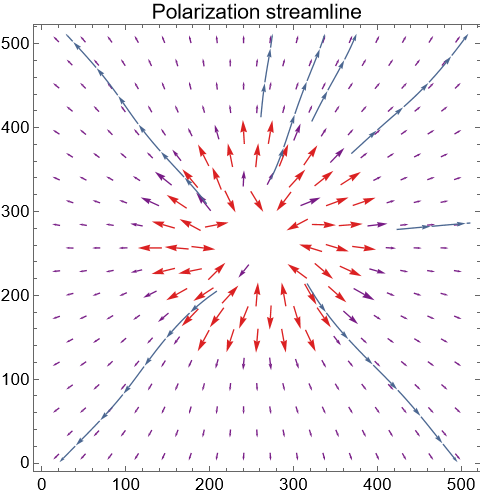}
\includegraphics[width=5cm,height=5cm]{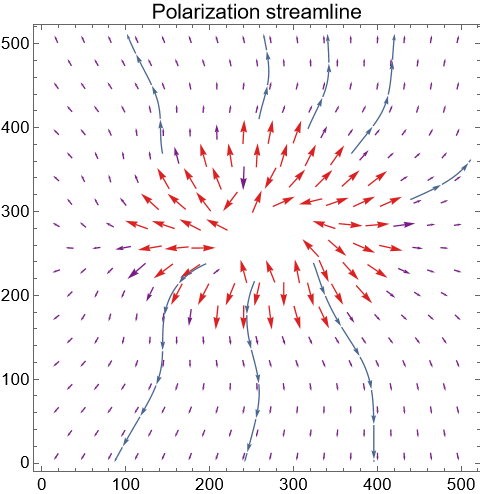}
\includegraphics[width=5cm,height=5cm]{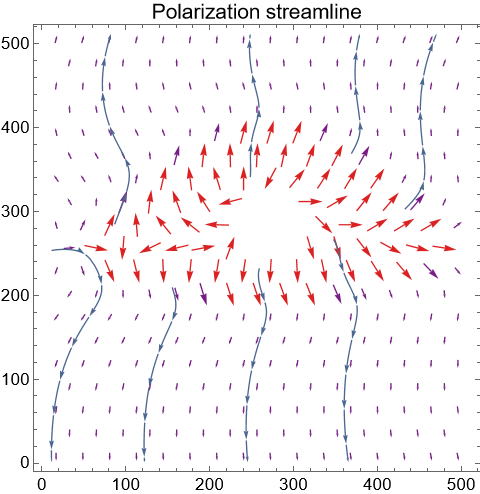}
\includegraphics[width=5cm,height=5cm]{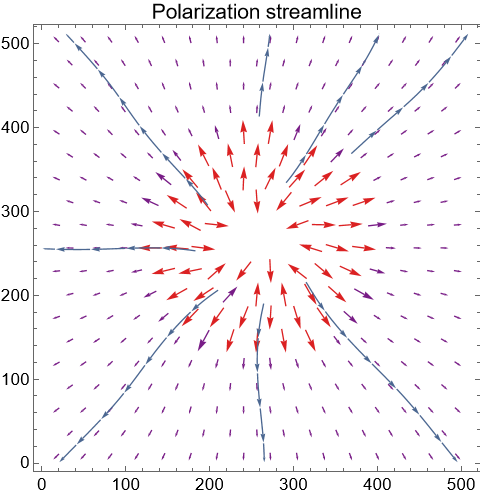}
\includegraphics[width=5cm,height=5cm]{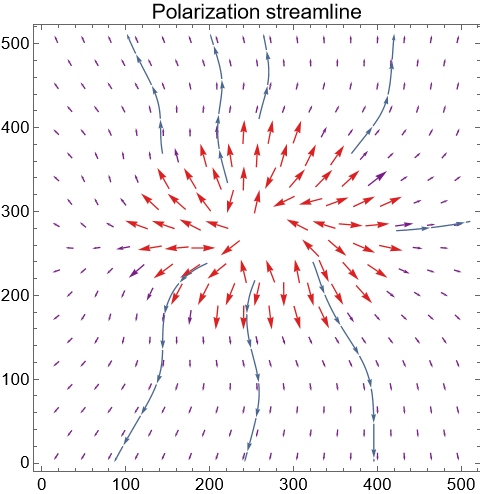}
\includegraphics[width=5cm,height=5cm]{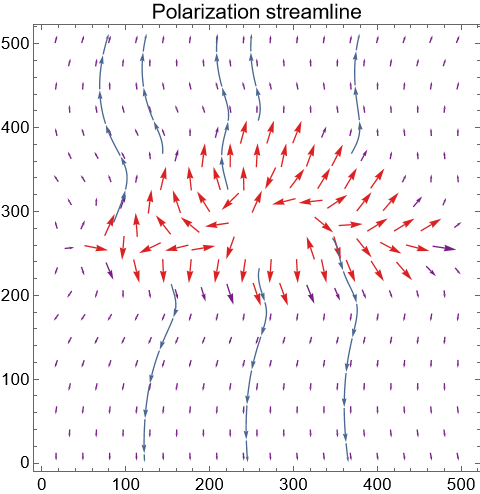}
\caption{\label{fig:12} The synchrotron radiation linearly polarized streamlines surrounded by a retrograde accretion disk around a Kerr-Newman black hole under different parameters.}
\end{figure*}

\begin{figure*}[t]
\centering
\includegraphics[width=5cm,height=5cm]{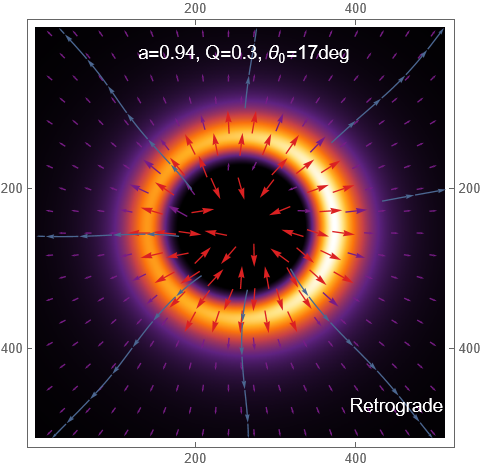}
\includegraphics[width=5cm,height=5cm]{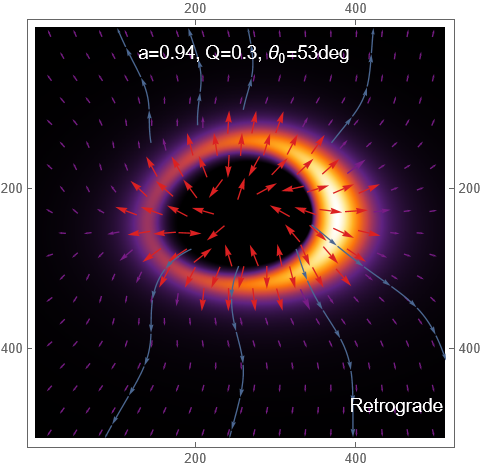}
\includegraphics[width=5cm,height=5cm]{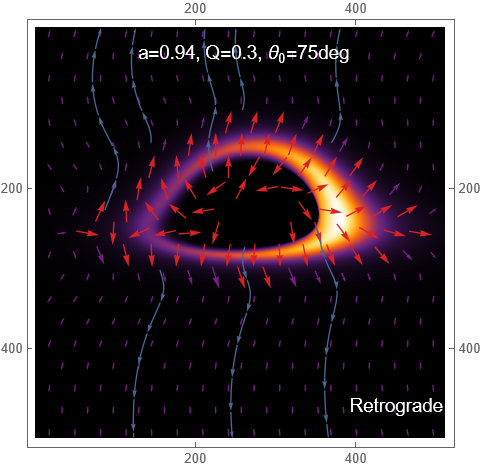}
\includegraphics[width=5cm,height=5cm]{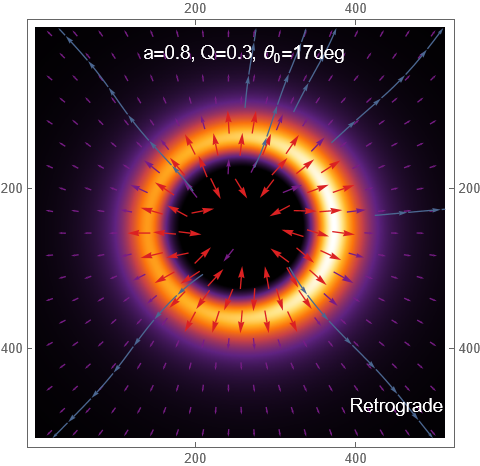}
\includegraphics[width=5cm,height=5cm]{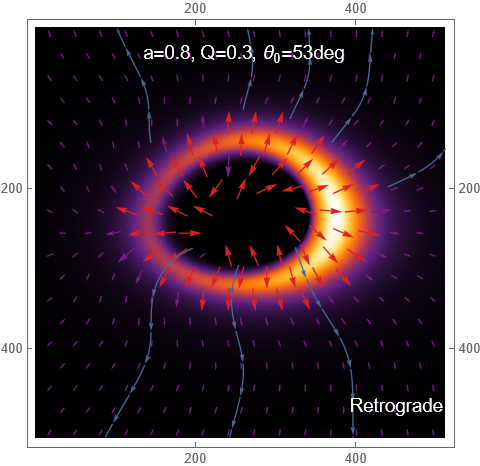}
\includegraphics[width=5cm,height=5cm]{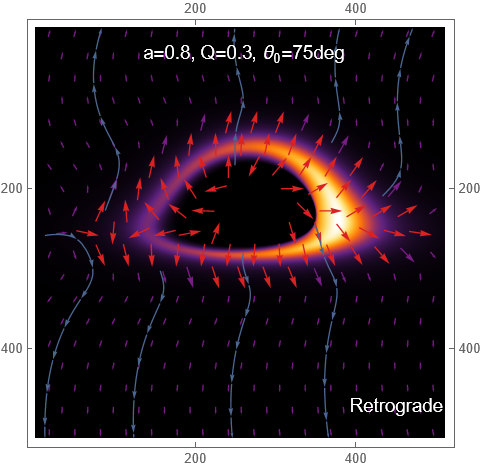}
\includegraphics[width=5cm,height=5cm]{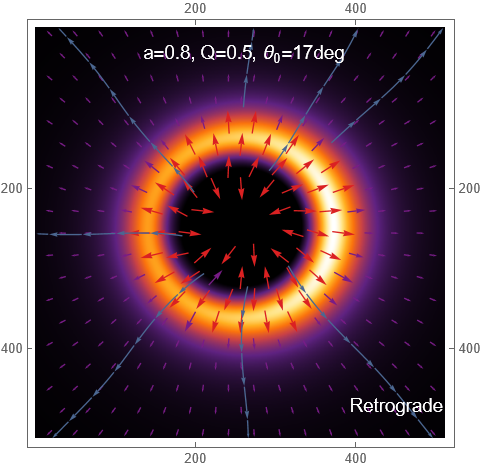}
\includegraphics[width=5cm,height=5cm]{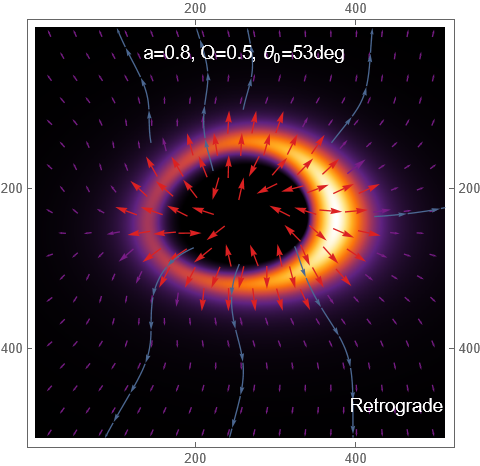}
\includegraphics[width=5cm,height=5cm]{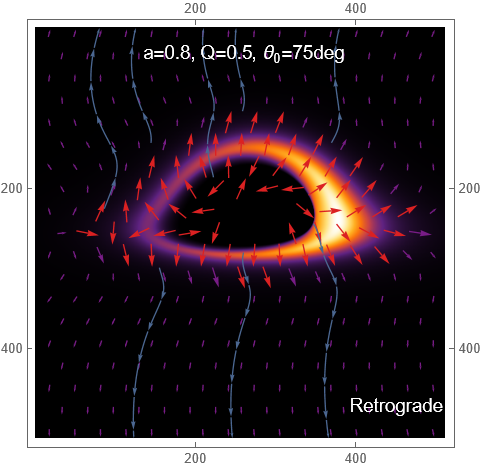}
\caption{\label{fig:13} Synchrotron radiation brightness and polarization distribution of Kerr-Newman black hole on retrograde disk.}
\end{figure*}

Next, we consider the case of a retrograde accretion disk. As shown in Figures~\ref{fig:11}-\ref{fig:13}, in the case of a prograde accretion disk, the material in the accretion disk rotates in the same direction as the spin of the black hole, resulting in the magnetic field and polarization streamlines generally exhibiting a pattern consistent with the spin direction. Polarization streamlines typically show a symmetric distribution, especially at low inclination angles. In contrast, for a retrograde accretion disk, the material rotates in the opposite direction to the black hole's spin. This reverse rotation alters the direction of the magnetic field and the distribution of polarization streamlines, resulting in a distinct pattern compared to that of a prograde accretion disk. The polarization streamlines of a retrograde accretion disk typically exhibit a twist opposite to the spin direction, especially at higher observation angles, where the structure of the polarization streamlines becomes more asymmetric. For both prograde and retrograde accretion disks, higher spin leads to magnetic field lines being more concentrated near the black hole. Lower spin results in a more extensive distribution of material in the accretion disk, and the changes in the polarization streamlines in a retrograde accretion disk are relatively gentle compared to those in a prograde accretion disk. A smaller charge weakens the electromagnetic field effects around the black hole, with only minor changes in the polarization streamlines. Regardless of whether the accretion disk is prograde or retrograde, the polarization streamlines generally maintain a symmetric distribution, but the streamlines in a retrograde accretion disk exhibit a different directional pattern compared to those in a prograde accretion disk. As the charge increases, the electromagnetic effect is enhanced, particularly in the retrograde accretion disk. Due to the influence of the charge, the distortion of the polarization streamlines becomes more complex, reflecting the strong impact of the charge on radiation.

Overall, there are significant differences in the distribution of polarization streamlines between prograde and retrograde accretion disks. Due to the reverse rotation of matter, the direction of polarization streamline twisting in retrograde accretion disks is opposite to that in prograde accretion disks. This effect is particularly pronounced under higher spin and charge conditions, where the asymmetry in the polarization streamlines becomes more evident.

\begin{figure}[t]
\centering
\includegraphics[width=6cm,height=8cm]{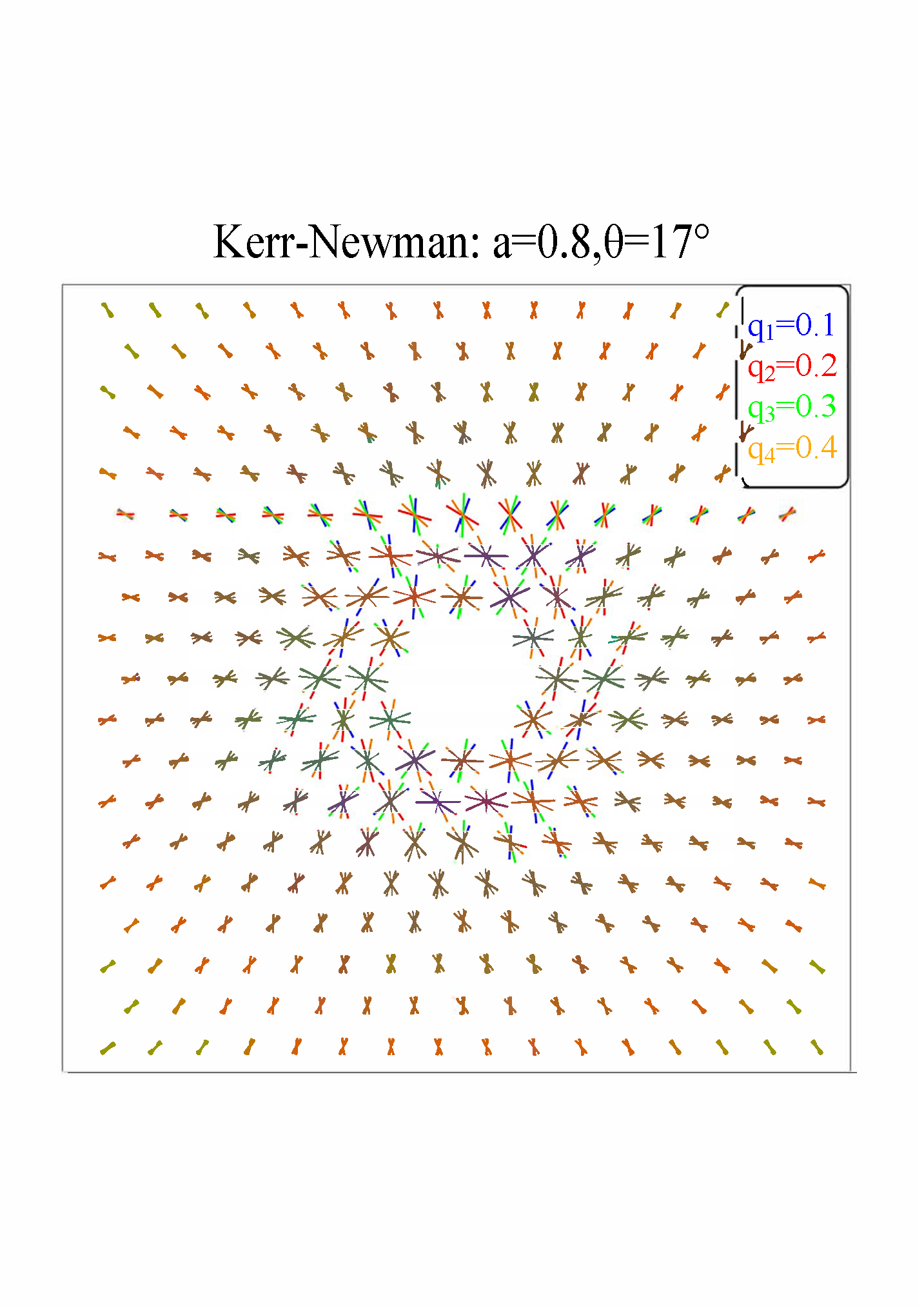}
\caption{\label{fig:14} EVPA patterns of the Kerr-Newman black hole for different charges. The blue, red, green, and orange vectors correspond to $Q=0.1$, $0.2$, $0.3$, and $0.4$, respectively.}
\end{figure}

Finally, we investigate the influence of black hole charge on synchrotron linear polarization. Figures~\ref{fig:14} shows the EVPA vector fields of a Kerr-Newman black hole for different charges at fixed spin $a=0.8$ and a low inclination of $\theta_0=17^\circ$. As the charge increases, the polarization vectors exhibit more prominent local rotations and displacements in the strong field region near the photon ring. The previously relatively ordered ring-like/near-radial pattern becomes progressively compressed and develops stronger asymmetric perturbations, whereas the large-scale morphology at larger radii remains only weakly affected. This trend indicates that charge modifies the spacetime structure in the strong gravity region and alters both the effective projection of the magnetic field and the parallel transport of the polarization vector, rendering the polarization topology increasingly sensitive to $q$. Consequently, systematic EVPA deviations around the photon-ring scale at low inclination may provide a potential observational diagnostic of charge effects.

\section{Conclusion and Discussion}
\label{sec:5}

In this analysis, we investigate the polarization radiation imaging of Kerr-Newman black holes by establishing a coupled ODE system that incorporates photon trajectory integration, polarization four-vector parallel transport, and charge-sensitive radiation emission models. We extend the traditional Walker-Penrose method to address its limitations when dealing with complex spacetimes. The Walker-Penrose method typically relies on specific symmetric structures and Killing tensors, which limit its applicability to analytically solvable spacetimes. By combining the photon orbit equation with the polarization parallel transport equation, we can evolve the photon trajectory and polarization state self-consistently without assuming specific symmetries. The advantage of this approach is that it can handle not only general axisymmetric spacetimes but also extend to non-axisymmetric geometries, thereby advancing our understanding of black hole radiation and polarization properties. This numerical framework allows us to simulate polarized radiation in more complex spacetime backgrounds, particularly considering the combined effects of black hole spin, accretion disk structure, and charge.

Through the analysis of the influence of black hole charge on polarization characteristics, we find that in the context of Kerr-Newman black holes, charge alters the photon propagation paths and polarization properties. As the black hole charge increases, the photon ring near the black hole gradually expands, and the ordered EVPA of the electric field is compressed, leading to changes in the topology of the polarization streamlines. Numerical simulations show that when the black hole spin is high and the charge is large, the expansion of the photon ring and the distortion of magnetic field lines become more pronounced, with polarization streamlines exhibiting more complex local distortions and rotations. These changes not only demonstrate the impact of black hole charge on the electromagnetic field and radiation but also provide new theoretical insights for future high-resolution observational tests of black hole charge. At high charge levels, the asymmetry in the polarization streamlines becomes more pronounced, particularly near the event horizon, where the electromagnetic field's influence on the radiation's polarization characteristics becomes stronger, further enhancing the effect of charge on radiation. Therefore, black hole charge is no longer an ignorable factor but an important physical quantity that significantly influences radiation polarization.

By introducing a tunable polarization source model, we provide a more flexible approach for modeling accretion disk radiation transfer. Traditional polarization models often rely on simple radiation source assumptions. By incorporating charge effects, we further explore how these effects generate complex nonlinear polarization characteristics in the radiation region around the black hole. More importantly, by introducing a self-consistent numerical framework, we successfully combine photon trajectories, polarization transport, and charge effects. This numerical framework provides a systematic ODE-based solution for black hole polarization imaging and enables further analysis of the asymmetry, polarity, and complexity of polarization in high-charge scenarios. A fully self-consistent treatment of charged black hole polarization would require coupling Kerr-Newman geometry to plasma dynamics and electromagnetic field evolution, which remains an open problem and is beyond the scope of the present study.


\acknowledgments

We greatly appreciate the insightful opinions provided by the referee, as well as the effective discussions between Professor Minyong Guo, Dr. Zhenyu Zhang, and Dr. Jiewei Huang. This work is supported by the National Natural Science Foundation of China (Grant No. 12133003, 12505060), Fapesq-PB of Brazil, the Fund Project of Chongqing Normal University (Grant Number: 24XLB033) and Chongqing Natural Science Foundation General Program (Grant No. CSTB2025NSCQ-GPX1019).


\end{document}